# Positivity-preserving hybrid DG/FV method for compressible Euler equations with stiff source terms

Zhen-Hua Jiang[1,a)], Xi Deng[2,3], Lin-Tao Huang[1], Chao Yan[1,a)], Feng Xiao[4], Jian Yu[1]

[1] College of Aeronautics Science and Engineering, Beijing University of Aeronautics and Astronautics, Beijing 100191, People's Republic of China

[2] Department of Chemical Engineering, Imperial College London, London SW7 2AZ, United Kingdom

[3] Department of Aeronautics, Imperial College London, London SW7 2AZ, United Kingdom

[4] Department of Mechanical Engineering, Tokyo Institute of Technology, 2-12-1(i6-29) Ookayama, Meguro-ku, Tokyo 152-8550, Japan

[a)] Corresponding author Email: jiangzhenhua@buaa.edu.cn (Zhen-Hua Jiang) and yanchao@buaa.edu.cn (Chao Yan)

## ABSTRACT

In numerical simulations of complex fluid dynamical problems, unphysical negative density or pressure may appear, causing blow-up of the computation. With the aim of obtaining positivity-preserving solutions with multi-scale resolution for solving strongly compressible flows, including hypersonic flows and stiff detonation waves, we present a positivity-preserving hybrid discontinuous Galerkin/finite volume (DG/FV) method. The approach is based on *a priori* and *a posteriori* computational methodology. The *a priori* computation utilizes relaxed boundary variation diminishing (BVD) algorithm to find troubled cells where the DG operators are replaced by the FV operators. The FV operators then deploy a hyperbolic tangent function in the reconstruction procedure to prevent unphysical values appearing in the flux evaluation. The *a posteriori* computation detects unphysical negative values in a simplified version of multidimensional optimal order detection (MOOD) paradigm and in the worst case applies the first-order FV Godunov scheme to guarantee the positivity of the solution. The *a priori* computation produces fewer oscillatory solutions, so that the *a posteriori* computation triggers less first-order evaluation. A technique of utilizing the tangent of hyperbola for interface capturing (THINC) with anti-diffusion effect, which is also referred to as the technique of adaptive reconstruction in this work, is suggested to reduce the numerical dissipation of the shock-capturing scheme. The current approaches retain the capability of the DG scheme to resolve small scales and the capability of the FV scheme to capture sharp discontinuities at the subgrid level. This favorable property is important to reproduce the correct position of detonation waves especially on the sparse grids. Numerical examples that are very challenging for high-order schemes are carried out to examine the performance of the approaches. The designed hybrid DG/FV scheme has obtained the positivity-preserving and oscillation-less solutions, as well as in some cases has achieved superior resolution in comparison with the fifth-order weighted essentially non-oscillatory (WENO) scheme using the same degrees of freedom (DOFs).

## 1 Introduction

There are a number of phenomena in physics and engineering that are described by hyperbolic systems of conservation laws. These include high-speed compressible flows, combustion and detonation problems. An important issue that arises when simulating such flows on the basis of hyperbolic conservation laws is the possible existence of discontinuous solutions even when the initial condition is smooth. Among the difficulties associated with the application of numerical methods to discontinuous solutions like shock waves is the appearance of unphysical oscillations and of unphysical values such as negative density and pressure. Besides the numerical challenges involved in obtaining non-oscillatory and positivity-preserving solutions, additional difficulties arise when the system of equations has a stiff source term. For example, numerical results for stiff detonation waves show that spurious oscillations become more severe when the spatial resolution is not high enough and that unphysical waves persist despite the use of a very small time step [1].

A loss of density or pressure positivity renders the system of equations nonconvex, leading to further difficulties in its numerical solution [2]. Much work has been done to improve numerical schemes in order to deal with such problems. There is a growing interest in designing high-order schemes that preserve positivity [3]. Successful high-order schemes, such as weighted essentially non-oscillatory (WENO) [4], discontinuous Galerkin (DG) [5] and flux reconstruction [6] schemes, have been constructed to preserve the positivity of density and pressure. In a series of works [7-10], the authors have established popular positivity-preserving approaches that maintain the designed order of accuracy of high order schemes. Other approaches introduce low order flux to the preservation of positivity [11-15]. For instance, the multidimensional optimal order detection (MOOD) [11], also known as *a posteriori* limiting paradigm, can be adapted as an effective procedure to preserve the positivity [16, 17].

In this work, we focus on rendering stability to numerical schemes for solving the reactive Euler equation system with stiff source terms. The source terms account for the effects of chemical reactions. The stiffness appears when the time scale of chemical reactions is significantly shorter than the flow dynamic time scale. Low-order numerical schemes can be used in solving the system of equations, but the results exhibit some deviations from experimental data [18]. Previous studies have attempted to employ high-order or high-resolution schemes to overcome this problem. The WENO schemes, see for instance [19, 20], and the DG schemes, see for instance [21-23], were utilized to improve the ability to capture detonation waves. The second-order schemes have also been developed for combustion and detonation problems by reducing numerical diffusion and enhancing the ability to resolve discontinuities [24, 25].

We have recently proposed a hybrid DG/FV method with subcell resolution [26]. The method has great advantages in resolving discontinuities and small-scale flow structures with high fidelity on coarser grids. Since the numerical viscosity in the shock-capturing scheme is critical for tracking the wave front in detonation simulations, we have been inspired to develop positivity-preserving hybrid DG/FV schemes to solve the reactive Euler equations. We believe that this work exhibits two differences compared to previous researches with regard to the solution of hyperbolic systems with stiff source terms [27]. First, most previous researches have concentrated on designing schemes with positivity-preserving property within one discretization framework. The present approaches preserve the positivity for the DG and FV schemes in a hybrid discretization framework. Second, the present approaches maintain the positivity and capture sharp discontinuities at the subgrid level, which is the key to reproduce the correct position of detonation waves on rather coarse meshes.

The remainder of the paper is organized as follows. Section 2 briefly introduces the governing equations.

Section 3 describes the detailed procedures to preserve the positivity for the hybrid DG/FV method. Numerical results are presented in Section 4, and some conclusions are drawn in Section 5.

## 2 Governing equations

The governing equations for the reactive Euler equations are written in the following form

$$\begin{aligned}
&\frac{\partial \rho}{\partial t}+\nabla\cdot(\rho \boldsymbol{u})=0,\\
&\frac{\partial \rho \boldsymbol{u}}{\partial t}+\nabla\cdot(\rho \boldsymbol{u}\boldsymbol{u}+p\underline{\boldsymbol{\delta}})=0,\\
&\frac{\partial \rho e}{\partial t}+\nabla\cdot[(\rho e+p)\boldsymbol{u}]=0,\\
&\frac{\partial \rho z_k}{\partial t}+\nabla\cdot\left(\rho z_k \boldsymbol{u}\right)=\omega_k,
\end{aligned} \tag{1}$$

where $t$ is the time, $\rho$ is the density, $\boldsymbol{u}$ is the velocity vector, $p$ is the static pressure, $e$ is the total energy per unit mass, $\underline{\boldsymbol{\delta}}$ is the Kronecker tensor, $z_k$ is the mass fraction of the $k$-th species, and $\omega_k$ is the mass production rate of the $k$-th species. The specific form of the chemical reaction source term $\omega_k$ is listed in the appendix. In addition, if we omit the last equation of Eq. (1), we obtain the governing equations for the general Euler equations, which are also studied in this work.

## 3 Positivity-preserving hybrid DG/FV method

### 3.1 Hybrid DG/FV methodology with subcell resolution

We rewrite Eq. (1) as

$$\frac{\partial \boldsymbol{Q}}{\partial t}+\nabla\cdot \boldsymbol{f}_c(\boldsymbol{Q})=\boldsymbol{s}, \tag{2}$$

where $\boldsymbol{Q}$ is the conservative variable vector, $\boldsymbol{f}_c(\boldsymbol{Q})$ is the inviscid flux tensor, and $\boldsymbol{s}$ is the source term. In this paper, the governing equation is solved by the so-called hybrid DG/FV methodology which carries out the DG discretization on the main element and the FV discretization on the subcell embedded in the main element, see also [28-30]. Here, we introduce the methodology by first formulating the DG discretization on the main element. The finite element solution $\boldsymbol{Q}_h$ is introduced as

$$\boldsymbol{Q}_h=\sum_{l=1}^{N(p)} v_{h,l} Q_l(t), \tag{3}$$

where $Q_l$ denotes the degree of freedom (DOF), $v_h$ is a piecewise-polynomial test function, and $N(p)$ is the number of modes. For the test function $v_h$, the Legendre basis is employed in the current work. By inserting $\boldsymbol{Q}_h$ in the governing equation and performing an integration by parts, the following weak formulation of Eq. (2) can be obtained as

$$\int_K v_h \frac{\partial \boldsymbol{Q}_h}{\partial t}\, d\boldsymbol{\Omega}+\int_{\partial K} v_h \boldsymbol{f}_c\cdot \boldsymbol{n}\, d\sigma-\int_K \nabla v_h\cdot \boldsymbol{f}_c(\boldsymbol{Q}_h)\, d\boldsymbol{\Omega}=\int_K v_h \boldsymbol{s}\, d\boldsymbol{\Omega}, \tag{4}$$

where $\boldsymbol{n}$ is the outward unit normal vector to the boundary. We utilize Gauss quadrature points for the domain and boundary integrals in Eq. (4). By assembling all the element contributions together, we can write Eq. (4) in the semi-discrete form as

$$M\frac{d\boldsymbol{Q}}{dt}+\boldsymbol{R}(\boldsymbol{Q})=0, \tag{5}$$

where we have the mass matrix $M$ and the residual vector $\boldsymbol{R}(\boldsymbol{Q})$ for the DG discretization.

In Eq. (3) we can see from the expression of DG solutions that multiple DOFs exist in the DG main element. It has been realized that the discontinuities can be captured within a single element by exploiting the multiscale information in the DG solution and the embedded FV subcell is one effective way to exploit this information. A representative set of FV subcells embedded in the fourth-order DG main element is illustrated in Fig. 1. The subcells are uniformly distributed and the number of the FV subcells is equal to the number of DG DOFs. On the embedded subcell the FV discretization is carried out. Take an example, the FV discretization on subcell $i, j$ shown in Fig. 1 is formulated as

$$\int_{ij}\frac{\partial \boldsymbol{q}}{\partial t}\,d\boldsymbol{\Omega}+\int_{ij}\nabla\cdot\boldsymbol{f}_c(\boldsymbol{q})\,d\boldsymbol{\Omega}=\int_{ij}\boldsymbol{s}\,d\boldsymbol{\Omega}, \tag{6}$$

where $\boldsymbol{q}$ indicates the conservative variable vector defined at the FV subcell. Based on the structured index in Fig. 1, we have

$$\frac{d\overline{\boldsymbol{q}}_{ij}(t)}{dt}+\hat{\boldsymbol{F}}_{c,\,ij+1/2}-\hat{\boldsymbol{F}}_{c,\,ij-1/2}=\Omega_{ij}\boldsymbol{s}(\overline{\boldsymbol{q}}_{ij}). \tag{7}$$

In Eq. (7), $\hat{\boldsymbol{F}}_{c,\,ij\pm1/2}$ is the convective flux at the subcell interface and $\overline{\boldsymbol{q}}_{ij}$ is the subcell average solution defined by

$$\overline{\boldsymbol{q}}_{ij}(t)=\frac{1}{\Omega_{ij}}\int_{x_{i-1/2}}^{x_{i+1/2}}\int_{y_{j-1/2}}^{y_{j+1/2}}\boldsymbol{q}(x,y,t)\,dydx. \tag{8}$$

Here $\Omega_{ij}$ denotes the subcell volume. $ij\pm1/2$ indicate both $x$ and $y$ components that include $i\pm1/2, j$ and $i, j\pm1/2$. It should be noted that the interface flux, including $\boldsymbol{f}_c\cdot\boldsymbol{n}$ in Eq. (4) and $\hat{\boldsymbol{F}}_{c,\,ij\pm1/2}$ in Eq. (7), is replaced by a monotone numerical Riemann flux function. In this paper, the local Lax–Friedrichs flux function [5] is employed for this task.

After the spatial discretization, the system of equations of the DG method and the system of equations of the FV method are both discretized in time by the third-order total variation diminishing (TVD) Runge–Kutta (RK) method [31]. For instance, the RK method to discretize the system of Eq. (5) is formulated as

$$\begin{aligned}
&\boldsymbol{Q}'=\boldsymbol{Q}^n-\Delta t M^{-1}\boldsymbol{R}(\boldsymbol{Q}),\\
&\boldsymbol{Q}''=\frac{3}{4}\boldsymbol{Q}^n+\frac{1}{4}\boldsymbol{Q}'-\frac{1}{4}\Delta t M^{-1}\boldsymbol{R}(\boldsymbol{Q}'),\\
&\boldsymbol{Q}^{n+1}=\frac{1}{3}\boldsymbol{Q}^n+\frac{2}{3}\boldsymbol{Q}''-\frac{2}{3}\Delta t M^{-1}\boldsymbol{R}(\boldsymbol{Q}''),
\end{aligned} \tag{9}$$

where $\boldsymbol{Q}^n$ and $\boldsymbol{Q}^{n+1}$ are the solutions at the present and next time step, while $\boldsymbol{Q}'$ and $\boldsymbol{Q}''$ denote the intermediate solutions.

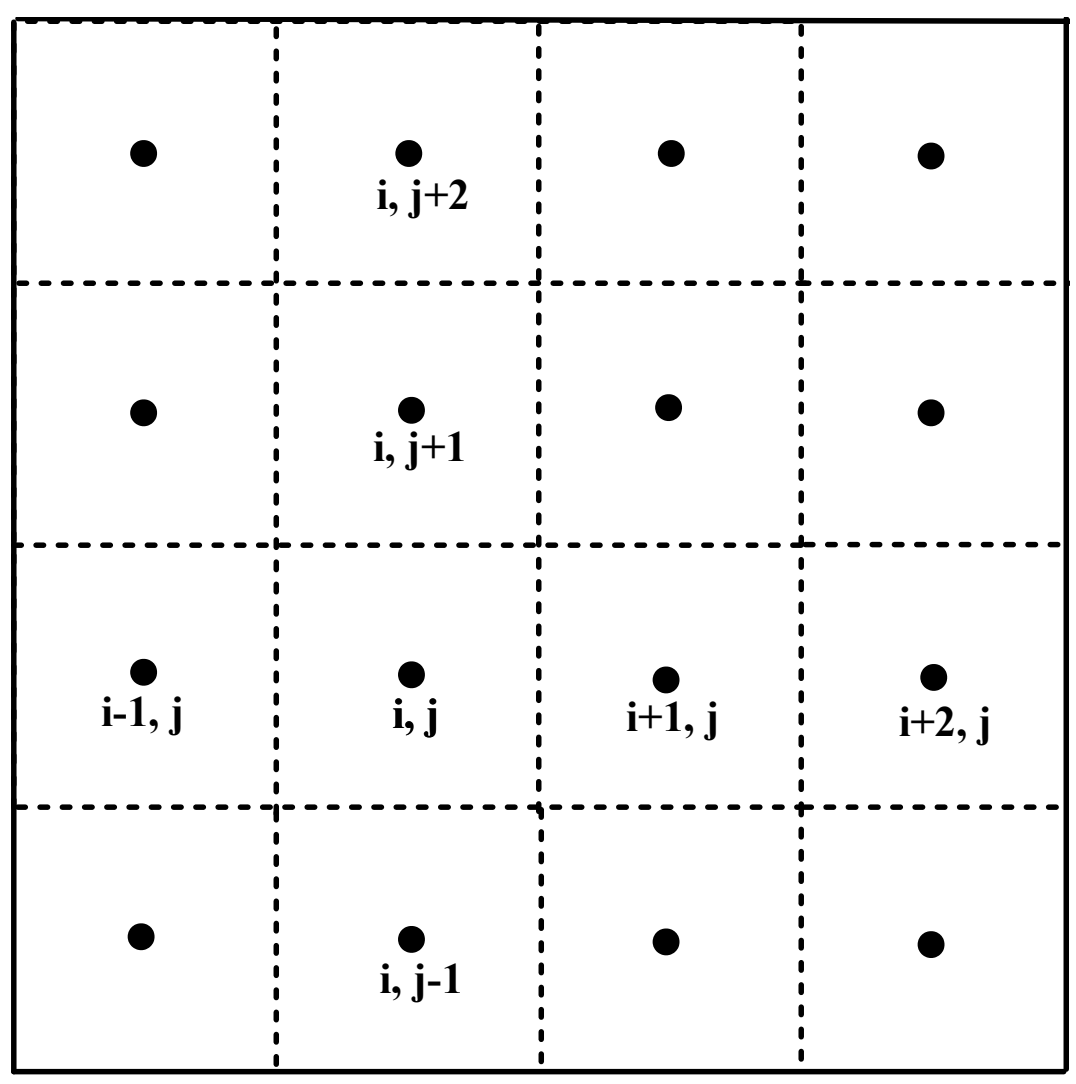

**Fig. 1** FV subcells (dashed line) and FV solutions (●) within the fourth-order DG main element (solid line).

### 3.2 Analyses of positivity constraints in the hybrid DG/FV methodology

In practice, the presence of a negative value leads to blow-ups of the numerical computation when the negative value appears under a radical sign. In a detailed analysis of the DG and FV discretization of the reactive Euler equations, we find two types of constraints with respect to the positivity of density and pressure: the first lies in the evaluation of the numerical flux, where the density and pressure at the left and right sides of the cell interface are required to be positive; the second lies in the evaluation of the updated cell average values, where the density and pressure of the volume-integrated average are required to be positive.

For the hybrid DG/FV method, it could have become more complex to preserve the positivity because there are two types of discretizations to deal with. Some researchers have tried to combine high order DG schemes and low order FV schemes at the subcell scale to ensure different properties on the numerical solution [32, 33]. The techniques such as the adaptive mesh refinement and the artificial viscosity method are also applicable in positivity-preserving DG-type methods [34, 35]. Nevertheless, the approaches presented in this work can make the preservation of positivity easier for the hybrid DG/FV method. The basic idea is to preserve the positive density and pressure of the volume-integrated subcell average for the FV computation, while preventing the negative values from being used for both the FV and the DG computations. Based on this idea, we design a strategy to obtain the positivity-preserving DG/FV method. A sketch illustrating the general procedure of the strategy is shown in Fig. 2. The designed positivity-preserving strategy indicated by the dashed lines consists of *a priori* detection/computation and *a posteriori* detection/computation. As can be seen from Fig. 2, the *a priori* and *a posteriori* detection/computation almost only target the process of the FV computation on the subcell. Also if we omit the positivity-preserving strategy that is indicated by the dashed lines in Fig. 2, the method reverts to the general procedure of the hybrid DG/FV method.

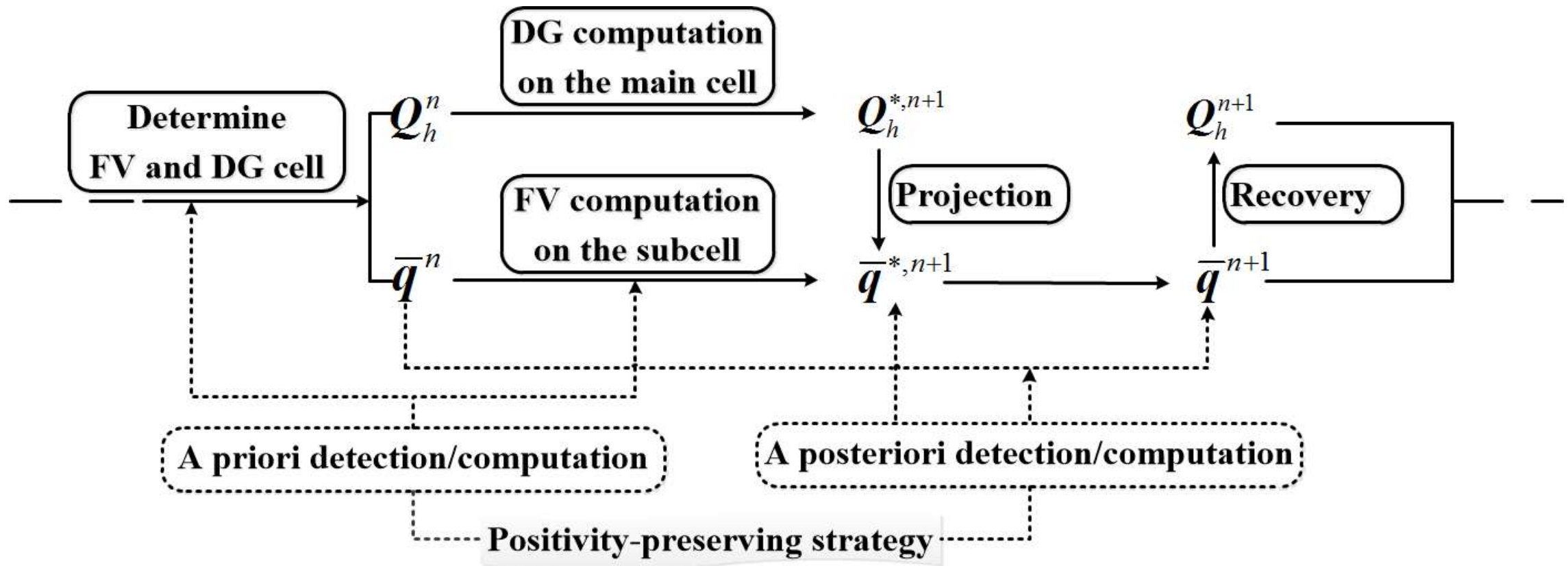

**Fig. 2** Sketch of the basic positivity-preserving strategy for the hybrid DG/FV methodology. $*$ indicates the candidate solution.

### 3.3 Preserving the positivity with *a priori* and *a posteriori* detection/computation

We now present the details of the designed positivity-preserving strategy. To keep the description simple, we consider each explicit substep discretization of the TVD RK method.

First, we define the subcell data representation used in this study, which is related to the projection and recovery process shown in Fig. 2. The projection process obtains the FV solution through the $L^2$ projection of the DG solution onto the subcell. Therefore, the FV solution of subcell $i, j$ is related to the DG solution through the projection process by

$$\overline{\boldsymbol{q}}_{ij} = \frac{1}{\Omega_{ij}} \int_{x_{i-1/2}}^{x_{i+1/2}} \int_{y_{j-1/2}}^{y_{j+1/2}} \boldsymbol{Q}_h(x, y, t)\, dydx. \tag{10}$$

One can see that the projection of the DG solution is consistent with the definition of the FV solution in Eq. (8). The number of Gaussian quadrature points utilized in the domain integration of subcell $i, j$ in Eq. (10) is the same as the number of Gaussian quadrature points used in the domain integration of main element $K$ in Eq. (4).

On the other hand, the recovery process involves obtaining the DG solution through the reverse process of the projection, i.e. utilizing the FV subcell solutions to obtain the DOFs of the DG solution. For the fourth-order DG element shown in Fig. 1, we insert Eq. (3) into Eq. (10) and list all the FV subcell solutions as

$$\begin{pmatrix} \overline{\boldsymbol{q}}_{i-1,j-1} \\ \cdot \\ \cdot \\ \cdot \\ \overline{\boldsymbol{q}}_{i+2,j+2} \end{pmatrix} = M_{16\times16} \begin{pmatrix} Q_1 \\ \cdot \\ \cdot \\ \cdot \\ Q_{16} \end{pmatrix}. \tag{11}$$

In this work, the number of FV subcells is equal to the number of DG DOFs. As a consequence, we have the matrices of 16×16 ($M_{16\times16}$) for both the projection and the recovery process of the fourth-order DG/FV scheme. The two matrices are the inverse matrices, which can be computed and stored before the iteration.

The cell-average definition of Eq. (10) is useful in preserving the conservation between the FV subcell solutions and DG main element solutions. From Eq. (10), we can infer that the main-element-averaged DG solution is equal to the summation of all FV subcell solutions. Hence, if the positivity of all FV subcell solutions is preserved, the positivity of the main-element-averaged DG solution is guaranteed.

According to the analysis in Sec. 3.2, there are two types of positivity constraints. We begin to design a strategy to satisfy the first type of positivity constraint. Given a positive FV solution $\overline{\boldsymbol{q}}^n$ on each subcell, we apply a reconstruction scheme that reconstructs the positive density and pressure on the left and right sides of the subcell interface. For this task, the tangent of hyperbola for interface capturing (THINC) function is chosen as the reconstruction scheme. The THINC [36, 37] scheme was originally proposed for multiphase simulations to preserve the boundedness of the volume of fluid (VOF) function, which requires the solution to be bounded by 0 and 1. The THINC scheme constructs the reconstruction function as

$$\boldsymbol{q}_i(x) = \overline{\boldsymbol{q}}_{\min} + \frac{\overline{\boldsymbol{q}}_{\max}}{2}\left\{1+\theta\tanh\left[\beta\left(\frac{x-x_{i-1/2}}{x_{i+1/2}-x_{i-1/2}}-\tilde{x}_i\right)\right]\right\}, \tag{12}$$

where $\overline{\boldsymbol{q}}_{\min}=\min(\overline{\boldsymbol{q}}_{i-1}, \overline{\boldsymbol{q}}_{i+1})$, $\overline{\boldsymbol{q}}_{\max}=\max(\overline{\boldsymbol{q}}_{i-1}, \overline{\boldsymbol{q}}_{i+1})-\overline{\boldsymbol{q}}_{\min}$ and $\theta=\mathrm{sgn}(\overline{\boldsymbol{q}}_{i+1}-\overline{\boldsymbol{q}}_{i-1})$. Here, we describe the THINC function in one dimension and hence drop the subscript *j*. In Eq. (12), the parameter $\beta$ controls the jump thickness of the THINC function and $\tilde{x}_i$ is computed from

$$\overline{\boldsymbol{q}}_i = \frac{1}{\Delta x}\int_{x_{i-1/2}}^{x_{i+1/2}} \boldsymbol{q}_i(x)\,dx. \tag{13}$$

As a sigmoid function, the hyperbolic tangent provides a good fit to a step-like discontinuity and also preserves the boundedness of the reconstruction. The THINC function can also be reformulated in the following form [38, 39]

$$\boldsymbol{q}_{i+1/2} = \overline{\boldsymbol{q}}_i + \frac{1}{2}\varphi(r_{i+1/2})(\overline{\boldsymbol{q}}_{i+1}-\overline{\boldsymbol{q}}_i), \tag{14}$$

which is similar to the TVD scheme. In Eq. (14) $r_{i+1/2}$ represents the ratio of upwind to central gradients and $\varphi(r_{i+1/2})$ is a nonlinear limiter function expressed as

$$\varphi(r_{i+1/2}) = 1-r_{i+1/2}+(1+r_{i+1/2})\frac{\tanh\beta+A}{1+A\tanh\beta}, \tag{15}$$

where

$$A=\frac{B/\cosh\beta-1}{\tanh\beta},\quad B=\exp[\theta\beta(2C-1)],\quad C=\frac{\overline{q}_i-\overline{q}_{\min}+\varepsilon}{\overline{q}_{\max}+\varepsilon},\quad \varepsilon=10^{-20}. \tag{16}$$

Here, $q$ denotes each component of the vector $\boldsymbol{q}$. The ad hoc parameter $\beta$ is set to be 1.1, which makes the THINC scheme have similar but slightly better performance than the TVD scheme with the van Leer limiter [24]. This part of the strategy applies the THINC scheme in the FV reconstruction process, therefore it belongs to the *a priori* computation acting on the FV computation on the subcell, as shown in Fig. 2.

Even with the THINC function adopted as the reconstruction scheme, the resulting method is still not positivity-preserving because the candidate solution $\overline{\boldsymbol{q}}^{*,n+1}$ in Fig. 2 can still become negative. This leads to the need for a positivity-preserving strategy with respect to the second type of positivity constraint, which maintains the positivity of subcell-averaged solutions. To be specific, we maintain the positivity of the FV solution $\overline{\boldsymbol{q}}^{n+1}$ in Fig. 2. In order to do so we apply the *a posteriori* detection acting on the candidate solution $\overline{\boldsymbol{q}}^{*,n+1}$ and apply the *a posteriori* computation acting on the process of obtaining $\overline{\boldsymbol{q}}^{n+1}$ from $\overline{\boldsymbol{q}}^{n}$, as can be seen in Fig. 2. The detailed procedure for this part of the strategy is summarized in pointwise form as follows:

1. Obtain $\overline{\boldsymbol{q}}^{*,n+1}$ either by the FV computation on the subcell for the FV cell or through the projection of the DG candidate solution $\boldsymbol{Q}_h^{*,n+1}$ for the DG element, as displayed in Fig. 2.
2. Check $\overline{\boldsymbol{q}}^{*,n+1}$ for the negative density or pressure. Mark the subcell that has the negative density or pressure as subcell $\overline{\boldsymbol{q}}^{*,n+1}_{\text{negative}}$ and mark the neighboring subcell that has the positive density and pressure but shares a common face with subcell $\overline{\boldsymbol{q}}^{*,n+1}_{\text{negative}}$ as subcell $\overline{\boldsymbol{q}}^{*,n+1}_{\text{neighbor}}$.
3. For subcell $\overline{\boldsymbol{q}}^{*,n+1}_{\text{negative}}$, use the last-step solution $\overline{\boldsymbol{q}}^{n}$ and perform the first-order Godunov scheme to recompute the residual. Store the recomputed first-order flux for the common faces that are shared with subcell $\overline{\boldsymbol{q}}^{*,n+1}_{\text{neighbor}}$.

4. For subcell $\overline{\boldsymbol{q}}^{*,n+1}_{\text{neighbor}}$, reassemble the subcell residual by using the first-order flux obtained in step 3 for the common faces and the previously computed flux in obtaining $\overline{\boldsymbol{q}}^{*,n+1}$ from $\overline{\boldsymbol{q}}^{n}$ for the rest faces.
5. Obtain the new solutions $\overline{\boldsymbol{q}}^{\text{new},n+1}_{\text{negative}}$ and $\overline{\boldsymbol{q}}^{\text{new},n+1}_{\text{neighbor}}$ using the recomputed residual and the reassembled residual for subcell $\overline{\boldsymbol{q}}^{*,n+1}_{\text{negative}}$ and subcell $\overline{\boldsymbol{q}}^{*,n+1}_{\text{neighbor}}$, respectively.
6. Return to step 2 and this time only check $\overline{\boldsymbol{q}}^{\text{new},n+1}_{\text{neighbor}}$ for the negative density or pressure. Repeat step 2 to step 6 until no negative density or pressure is detected for $\overline{\boldsymbol{q}}^{*,n+1}$.
7. Set $\overline{\boldsymbol{q}}^{n+1}=\overline{\boldsymbol{q}}^{*,n+1}$, perform the recovery process to obtain the DG solution $\boldsymbol{Q}_h^{n+1}$, and then continue the computation as displayed in Fig. 2.

A detail of the above procedure is noted that in step 4 the flux of the faces that are not shared with subcell $\overline{\boldsymbol{q}}^{*,n+1}_{\text{negative}}$ is required to compute $\overline{\boldsymbol{q}}^{*,n+1}$ from $\overline{\boldsymbol{q}}^{n}$. Therefore when subcell $\overline{\boldsymbol{q}}^{*,n+1}_{\text{neighbor}}$ appears in the DG element, the FV computation has to be re-performed for subcell $\overline{\boldsymbol{q}}^{*,n+1}_{\text{neighbor}}$ to obtain the flux because in the original DG/FV method there is no FV flux for subcells within the DG element.

Another detail of the above procedure concerns the evaluation of the flux on the interface connecting the DG element and the FV cell. In this work, we set the flux on such common interfaces as the flux computed by the FV method. To be specific, we first define a Lagrange polynomial of the DG flux based on the Gauss integration points on the common interface, and then set the integrals of the DG flux on each sub-interface of the FV subcell equal to the FV flux on the corresponding sub-interface. Since the number of the unknown DG integration point flux is equal to the number of sub-interfaces, the DG integration point flux on the common interface can be determined.

The above procedure detects the candidate solution $\overline{\boldsymbol{q}}^{*,n+1}$ and calculates $\overline{\boldsymbol{q}}^{n+1}$ by solving the governing equations using $\overline{\boldsymbol{q}}^{n}$. Therefore the procedure belongs to an *a posteriori* detection/computation method shown in Fig. 2. The idea here is very similar to the multidimensional optimal order detection (MOOD) method [11]. The MOOD has been applied as an *a posteriori* detection paradigm for designing shock-capturing schemes in the context of various discretization methods, see for instance [16, 29, 40-42]. The strategy presented here can be considered as a simplified version of MOOD, because it only applies one loop using the first-order scheme for cells with unphysical negative values. The *a posteriori* computation is in fact only activated in few cells where extremely complex computational condition appears, as will be demonstrated in the following numerical experiment.

### 3.4 Indication strategy for the positivity-preserving hybrid DG/FV method

In the last subsection, we have described the three parts of current positivity-preserving strategy, including the *a priori* computation, the *a posteriori* detection and the *a posteriori* computation. The only left part of the strategy is the *a priori* detection that will be described in this subsection. As can be seen in Fig. 2, the *a priori* detection acts on the process of determining the FV and DG cell and the process of the FV computation on the subcell.

When the *a priori* detection acts on the process of determining the FV and DG cell, the *a priori* detection actually belongs to a portion of the strategies known as the troubled-cell indicator. The troubled-cell indicator is often required for the construction of the so-called hybrid scheme. This study extends the troubled-cell indicator proposed in [26] to the reactive Euler equations. The main steps of the troubled-cell indicator are catalogued as follows.

Firstly, we perform a high-order reconstruction and a low-order reconstruction to obtain the variables on the left and right sides of the subcell interface. Two sets of interface variables are indicated by superscripts

<high-order> and <low-order> in Eqs. (17)–(20) below. Also, in Eqs. (17)-(18) $\boldsymbol{q}_{i\pm1/2}^{L,R}$ denote the variables on the left and right sides of interfaces in the $x$ direction for subcell $i,j$. We calculate the total boundary variation (TBV) values in the $x$ direction for the high- and low- order reconstruction as

$$TBV_i^{\langle\text{high-order}\rangle} = \left|\boldsymbol{q}_{i-1/2}^{L,\langle\text{high-order}\rangle} - \boldsymbol{q}_{i-1/2}^{R,\langle\text{high-order}\rangle}\right| + \left|\boldsymbol{q}_{i+1/2}^{L,\langle\text{high-order}\rangle} - \boldsymbol{q}_{i+1/2}^{R,\langle\text{high-order}\rangle}\right|, \tag{17}$$

$$TBV_i^{\langle\text{low-order}\rangle} = \left|\boldsymbol{q}_{i-1/2}^{L,\langle\text{low-order}\rangle} - \boldsymbol{q}_{i-1/2}^{R,\langle\text{low-order}\rangle}\right| + \left|\boldsymbol{q}_{i+1/2}^{L,\langle\text{low-order}\rangle} - \boldsymbol{q}_{i+1/2}^{R,\langle\text{low-order}\rangle}\right|. \tag{18}$$

The same can be done to obtain $TBV_j^{\langle\text{high-order}\rangle}$ and $TBV_j^{\langle\text{low-order}\rangle}$ in the $y$ direction. Following more recent studies in [43], we choose the high-order (specifically seventh-order) linear reconstruction and the THINC reconstruction (specifically $\beta$=1.1) as the high- and low- order reconstruction in this work.

Secondly, we adopt a so-called relaxed boundary variation diminishing (BVD) condition [26, 44] to compare the $TBV^{\langle\text{high-order}\rangle}$ and $TBV^{\langle\text{low-order}\rangle}$ in the $x$ and $y$ directions as:

$$TBV_i^{\langle\text{high-order}\rangle} - TBV_i^{\langle\text{low-order}\rangle} > \max\left(\varepsilon_1, \varepsilon_2\left(\left|\boldsymbol{q}_{i-1/2}^{L,\langle\text{low-order}\rangle}\right| + \left|\boldsymbol{q}_{i-1/2}^{R,\langle\text{low-order}\rangle}\right| + \left|\boldsymbol{q}_{i+1/2}^{L,\langle\text{low-order}\rangle}\right| + \left|\boldsymbol{q}_{i+1/2}^{R,\langle\text{low-order}\rangle}\right|\right)\right), \tag{19}$$

$$TBV_j^{\langle\text{high-order}\rangle} - TBV_j^{\langle\text{low-order}\rangle} > \max\left(\varepsilon_1, \varepsilon_2\left(\left|\boldsymbol{q}_{j-1/2}^{L,\langle\text{low-order}\rangle}\right| + \left|\boldsymbol{q}_{j-1/2}^{R,\langle\text{low-order}\rangle}\right| + \left|\boldsymbol{q}_{j+1/2}^{L,\langle\text{low-order}\rangle}\right| + \left|\boldsymbol{q}_{j+1/2}^{R,\langle\text{low-order}\rangle}\right|\right)\right). \tag{20}$$

We set $\varepsilon_1=10^{-6}$ and $\varepsilon_2=10^{-4}$ in this work. Then, for subcell $i,j$, we use $Nrec_{ij}$ to count the total number of reconstructions that satisfy either Eq. (19) or Eq. (20). Having obtained $Nrec$ for all subcells in main element $K$, we then use $Nsub_K$ to count the total number of subcells that have $Nrec_{lm}>0$, $l=i-1, \ldots, i+2$ and $m=j-1, \ldots, j+2$.

Thirdly, main element $K$ is determined as a FV cell if any $Nrec_{lm}>1$ for $l=i-1, \ldots, i+2$ and $m=j-1, \ldots, j+2$, or $Nsub_K>N(p)/2$, where $N(p)$ is the total number of subcells in main element $K$. Main element $K$ is also determined as a FV cell if the positivity of density or pressure is violated at any Gaussian quadrature point of volume and face integrations in the DG computation. Otherwise, main element $K$ is determined as a DG element.

On the other hand, when the *a priori* detection acts on the process of the FV computation on the subcell, the *a priori* detection checks the positivity of the reconstructed variables of density and pressure on the left and right sides of the subcell interface for the FV cell. Sec. 3.3 has adopted the THINC reconstruction ($\beta$=1.1) as the *a priori* computation to get the positive density and pressure on the left and right sides of the subcell interface. However, as will be described in Sec. 3.5, this work in fact hybridize the THINC scheme ($\beta$=1.1) with other scheme in order to reduce the numerical dissipation. This hybridization of different reconstruction schemes, or the so-called adaptive reconstruction in Sec. 3.5, cannot guarantee the positivity of the reconstructed density and pressure. Therefore, we still check the positivity of the reconstructed variables on the subcell interface. If the positivity of the density and pressure reconstructed by the adaptive reconstruction is violated, we go back to reconstruct the interface variables using the THINC scheme ($\beta$=1.1). If the positivity of the density and pressure reconstructed by the THINC scheme ($\beta$=1.1) is still violated, although this situation has not occurred in the cases considered in this work, one can use the first-order reconstruction since the FV subcell solution is positive guaranteed by the method in Sec. 3.3.

The positivity-preserving hybrid DG/FV method is now complete.

### 3.5 Extension to the reactive Euler equations with the adaptive reconstruction

The positivity-preserving DG/FV method presented above is essentially a shock-capturing scheme that still produces a few transition points in the shock. For stiff cases like the detonation problem, these transition points are responsible for incorrect propagation speeds of discontinuities [1]. Furthermore, the numerical dissipation of the shock-capturing scheme, although at the subgrid level, still causes a certain loss of accuracy of the hybrid

DG/FV method [26].

Here, we provide a technique to reduce the numerical dissipation for the positivity-preserving scheme. The basic idea is to hybridize the THINC scheme ($\beta$=1.1) with other schemes that have anti-diffusion effects on numerical solutions. According to [24], the THINC scheme with larger $\beta$ generates anti-diffusion effects. Therefore, in the FV reconstruction process we adaptively choose the THINC scheme with $\beta$=1.1 and the THINC scheme with larger $\beta$. To be specific, firstly using Eq. (12), we perform the THINC reconstruction with small $\beta_s$=1.1 and large $\beta_l$=1.6 [24]. The two reconstructions are thus denoted by $\boldsymbol{q}^{<\beta_s>}(x)$ and $\boldsymbol{q}^{<\beta_l>}(x)$ to reconstruct subcell interface variables $\boldsymbol{q}_{i\pm1/2}^{L,R<\beta_s>}$ and $\boldsymbol{q}_{i\pm1/2}^{L,R<\beta_l>}$, respectively. Then, we use equations similar to Eqs. (17)-(18) to calculate $TBV_i^{<\beta_s>}$ and $TBV_i^{<\beta_l>}$. Lastly, we determine the final reconstruction for subcell $i$ using the BVD selection algorithm

$$\boldsymbol{q}_i^{<\text{Final}>}(x) = \begin{cases} \boldsymbol{q}_i^{<\beta_s>}(x) & \text{if} \;\; TBV_i^{<\beta_l>} > TBV_i^{<\beta_s>}, \\ \boldsymbol{q}_i^{<\beta_l>}(x) & \text{otherwise}. \end{cases} \tag{21}$$

The above procedure of adaptively choosing the THINC scheme with $\beta_s$=1.1 and $\beta_l$=1.6 is also referred to as the adaptive reconstruction in this work. An illustration of the connection between the DG/FV method and the adaptive reconstruction is presented in Fig. 3, where a one-dimensional reconstruction is shown for simplicity. Let us assume the discontinuity exists in subcell $i$. Firstly, the seventh-order linear reconstruction and THINC reconstruction ($\beta$=$\beta_s$=1.1) are performed for subcells $i-5$, …, $i$+6. By the indication strategy presented in Sec. 3.4, let us assume the center element is identified as the FV cell while the other two elements identified as the DG cells, as shown in Fig. 3. Then, the THINC reconstruction ($\beta$=$\beta_l$=1.6) is performed for subcells $i-1$, …, $i$+2. By the BVD selection algorithm of Eq. (21), let us assume only subcell $i$ has the reconstruction $\boldsymbol{q}_i^{<\beta_s>}(x)$ changed into $\boldsymbol{q}_i^{<\beta_l>}(x)$ while the rest subcells not, as shown in Fig. 3. Therefore the goal to obtain the sharper transition profile for the discontinuity at the sub-grid level is achieved.

The adaptive reconstruction presented above can already give favorable results for the cases considered in this work. Nevertheless, it is still possible to add other reconstruction functions to the reconstruction process to achieve an even less diffusive method. For instance, by adding the fifth-order linear reconstruction and using a two-staged BVD selection algorithm [43, 26], the numerical dissipation of the scheme can be further reduced, as will be demonstrated in some of the following numerical tests. Here we do not give the details of adding the fifth-order reconstruction and using a two-staged BVD selection algorithm, because they have been presented in detail in the FV [43] and the DG [26] framework. Unless otherwise specified, the following numerical results are obtained by the adaptive reconstruction of hybridizing the THINC scheme ($\beta_s$=1.1) with the THINC scheme ($\beta_l$=1.6).

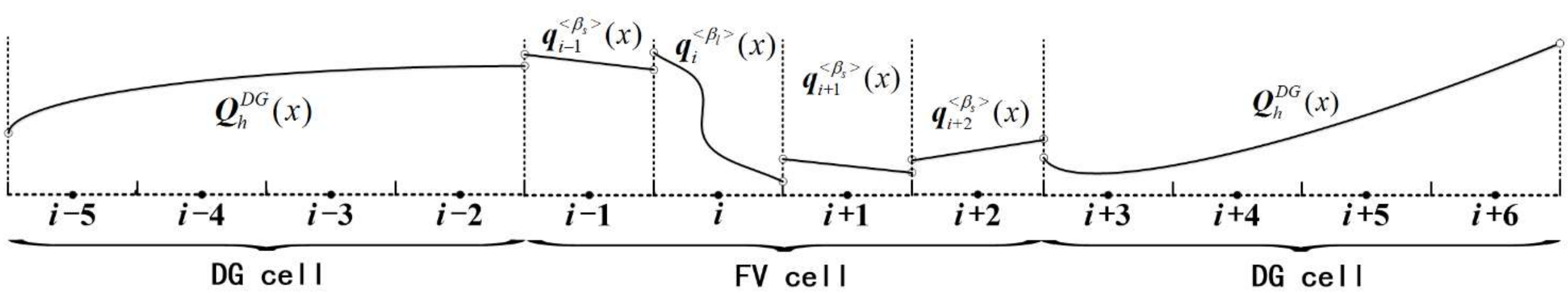

**Fig. 3** Illustration of the adaptive reconstruction in the hybrid DG/FV method.

### 3.6 Further discussion of the present positivity-preserving hybrid DG/FV method

In this subsection we would like to discuss some numerical properties as well as some implementation details about the present method.

Firstly, we use the density wave transport problem to show the accuracy properties of the proposed method. The exact solution of the problem is $\rho = 1 + 0.9999\sin\theta,\ u = 1,\ v = 1,\ p = 1,\ \theta = x + y - 2t$. The minimum density of the exact solution is 0.0001. The Euler equations are solved for this problem on the domain of $[0, 2\pi] \times [0, 2\pi]$ from the time $t$=0 to $t$=0.1. Table 1 lists the errors of the cell-averaged density for the proposed method using the 2nd order scheme, as well as the errors of the FV method that utilizes the same DOFs as the proposed method. The THINC with $\beta$=1.1 is used as shock-capturing scheme for the FV method. One can see that the present DG/FV overall maintains the designed 2nd-order accuracy, while the FV achieves the accuracy of less than the 2nd-order. We would like to point out that for all five meshes considered in Table 1 for the DG/FV method, the FV cells are activated in almost the whole computation process. The averaged number of FV cells during entire iteration steps are about 11, 20, 30, 19 and 15 for the meshes ranging from 5×5 to 80×80. For the DG/FV method, the 1st-order FV computation is also activated on several FV subcells during a period of time, while for the FV method, there are no any positivity-preserving strategy required for this test case. When we test the 3rd-order DG/FV method, the FV cells are not activated on most of meshes. In order to examine the accuracy, we enforce the fixed number FV cells for the 3rd-order DG/FV computation. The fixed number is the same as the average number of the 2nd-order method, which is 11, 20, 30, 19 and 15 on the corresponding mesh. The enforced FV cells are chosen randomly for each iteration. From Table 2, we see the convergence rates largely oscillate between the 2nd- and 3rd- order of accuracy.

It is our belief that the uniform high-order accuracy of the original DG method will be compromised since the low order FV scheme is utilized. Nevertheless, for cells with extremely complex conditions, the high-order accuracy of the method is not mainly concerned as long as the solution properties, like the positive density and pressure, are preserved on such cells. Similar point of view can also be seen in [17]. This work does not intend to achieve the formal high-order accuracy but mainly focuses on the positivity-preserving ability and the high-resolution property for solving the Euler equations with stiff source terms. We mention that the uniform high-order accuracy may be also realizable, if for instance the method in [3] is adopted in the current framework. It requires further investigation beyond the scope of this study.

**Table 1** Density errors and convergence rates for the 2nd order DG/FV method and the FV method

| Mesh size for DG/FV (P1) | DG/FV (P1) $L^2$ error | DG/FV (P1) Order | Mesh size for FV | FV $L^2$ error | FV Order |
|---|---|---|---|---|---|
| 5×5 | 1.75e-2 | — | 10×10 | 4.50e-2 | — |
| 10×10 | 4.78e-3 | 1.87 | 20×20 | 1.66e-2 | 1.44 |
| 20×20 | 1.42e-3 | 1.75 | 40×40 | 6.37e-3 | 1.38 |
| 40×40 | 2.49e-4 | 2.51 | 80×80 | 2.45e-3 | 1.38 |
| 80×80 | 5.84e-5 | 2.09 | 160×160 | 9.20e-4 | 1.41 |

**Table 2** Density errors and convergence rates for the 3rd order DG/FV method with fixed number of FV cells

| Mesh size for DG/FV (P2) | DG/FV (P2) $L^2$ error | DG/FV (P2) Order |
|---|---|---|
| 5×5 | 4.48e-3 | — |
| 10×10 | 1.47e-3 | 1.61 |
| 20×20 | 1.06e-4 | 3.79 |
| 40×40 | 1.71e-5 | 2.63 |
| 80×80 | 1.67e-6 | 3.35 |

Then we give some implementation details before we begin the following numerical experimentation. First,

the time step is determined on the sub-grid level in this work, and the CFL number is chosen as 0.3 for the following numerical test cases. Second, unless otherwise specified, this work adopts the primitive variables in solving the reactive Euler equations. Third, the reactant mass fraction $z$ tends to become negative in some cases. The present method chooses to simply set $z$ to be zero if $z<0$, especially when $z$ is used in the computation of the source term. Fourth, the current positivity-preserving strategy is applicable to arbitrary-order DG/FV schemes as well as to the unstructured meshes, although in the following we mainly test the fourth-order scheme on the rectangular mesh.

## 4 Numerical tests

### 4.1 Numerical results for the nonreactive Euler equations

In this subsection, we test the developed method for the nonreactive Euler equations. Two typical cases are chosen to examine the performance of the approaches. The characteristic variable is used in the reconstruction process for these two cases. The results are shown by using the subcell solutions.

#### 4.1.1 High Mach number flows

The first test case is the Mach 2000 problem, also investigated in [8, 45, 46], which is chosen to demonstrate the positivity-preserving ability of the method. The computational domain is a rectangle of $[0, 1]\times[-0.25, 0.25]$ initially full of the ambient gas with density of 0.5, velocity of 0 and pressure of 0.4127. The boundary conditions are outflow on the top, bottom and right edges. On the left boundary, a jet with a width of 0.1 located by $-0.05\leqslant y\leqslant 0.05$ enters. The jet has density of 5.0, velocity of 800 and pressure of 0.4127. For the rest of the left boundary, the density is 0.5, the velocity is 0 and the pressure is 0.4127. The specific heat $\gamma$ is set as 5/3 and the computation stops at $t$=0.001. Fig. 4 shows the density counters on $160\times80$ main elements, where the solutions on the subcell are illustrated. With a coarser mesh adopted here, the results are still comparable to those reported in Fig. 4.6 of [8]. In Fig. 4 we also show main elements containing subcells that have once activated the first-order FV computation during the whole computation process. These elements are mainly around and inside the jet. The lack of distribution symmetry of these elements has less effect on the symmetry of numerical solutions. We also count the number of subcells that activate the first-order FV computation in each iteration. We observe the largest subcell number is only 28 at time $t\approx2.2\times10^{-5}$ and for the rest of computational time the subcell number is often less than 4 in each iteration. These numerical observations are consistent with the analysis presented in Sec. 3.3.

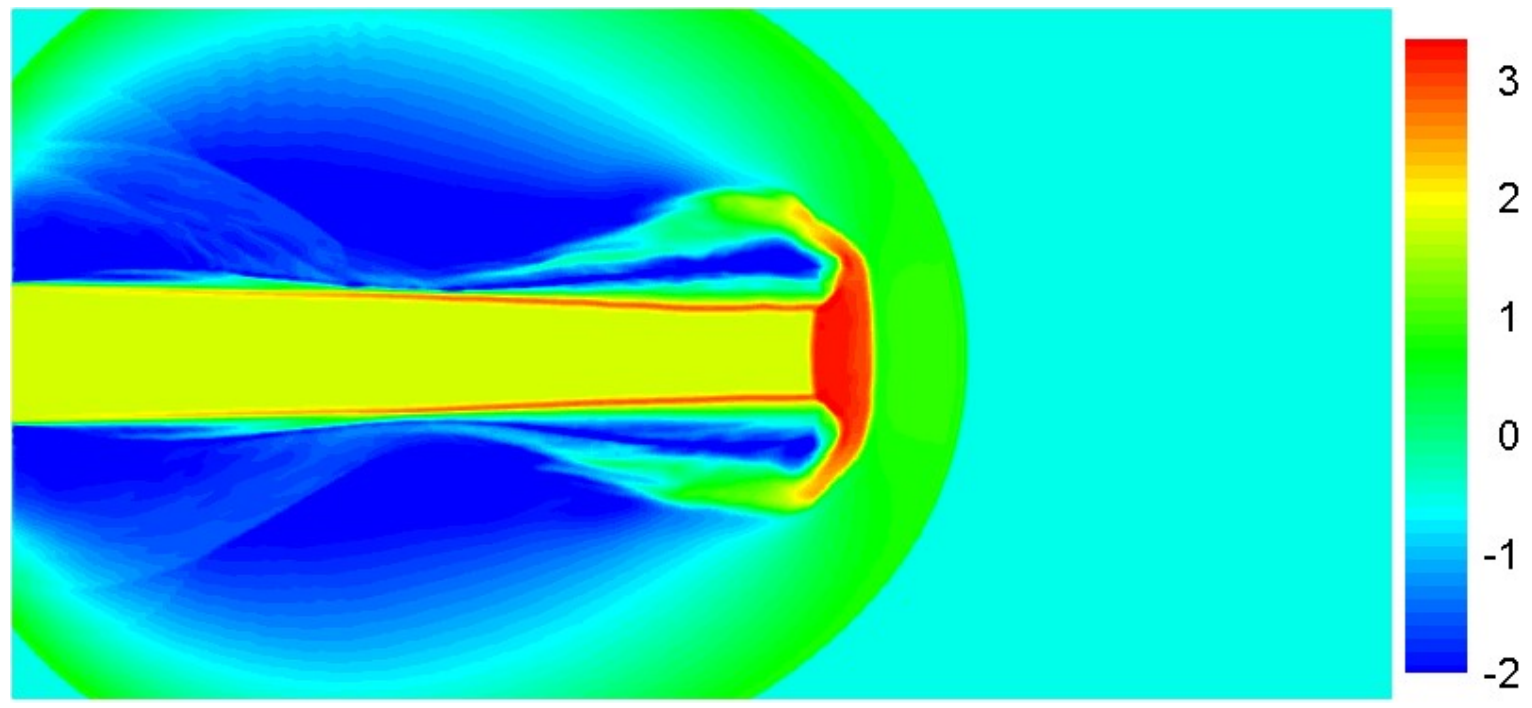

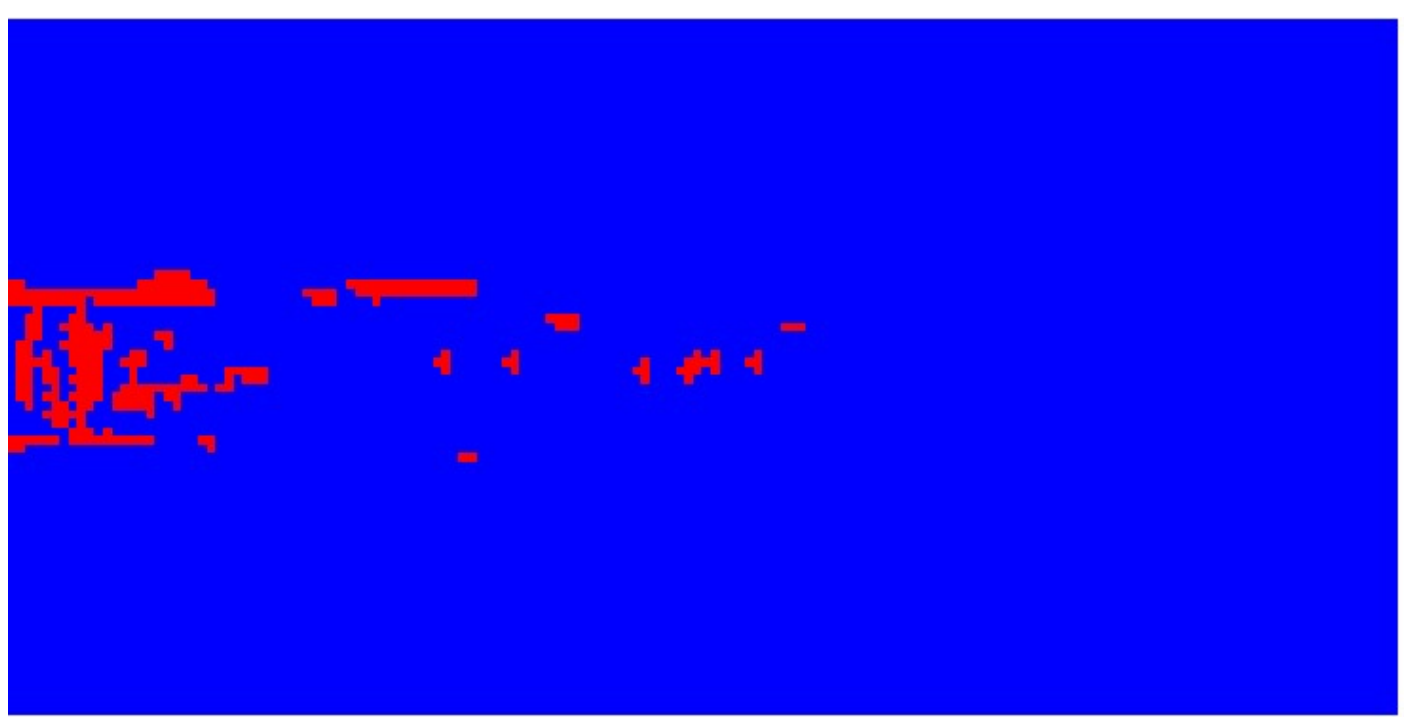

**Fig. 4** Mach 2000 jet. Upper: density solutions on the subcell. Scales are logarithmic; lower: red color elements are the main elements containing subcells that have once activated the first-order FV computation during the whole computation process.

### 4.1.2 Reflection and diffraction of a Mach 10 shock

The second case describes a Mach 10 shock passes an equilateral triangle, also known as Schardin's problem [47]. This case is selected because it has not only low density and pressure but also complicated structures due to Kelvin–Helmholtz instability. An illustration of the domain and the mesh is presented in Fig. 5. The quadrilateral meshes used in the computation have cell edge length around 1/80. Initially, a right-moving Mach 10 shock is placed at $x$=0.2. The preshock state is ($\rho$, $u$, $v$, $p$)=(1.4, 0.0, 0.0, 1.0). The inflow and outflow boundary conditions are applied for the left and right boundaries. The exact solution of the right-moving Mach 10 shock is adopted for the top boundary. The reflective boundary conditions are employed for the rest boundaries.

The density counters and the troubled cells at time $t$=0.245 are shown in Figs. 6-7. The results in the left figures are obtained by the present method that hybridizes the THINC scheme ($\beta_s$=1.1) with the THINC scheme ($\beta_l$=1.6). The results in the right figures are obtained by the method mentioned in Sec. 3.5, which can further reduce numerical dissipation by adding the fifth-order reconstruction and using a two-staged BVD selection algorithm. The density solutions on the subcell scale are again kept positive. As expected, the method of adding the fifth-order reconstruction has captured more delicate structures although more troubled cells are identified. In addition, the subcell resolution on discontinuity-capturing can be seen in Fig. 8, where the shock profile is captured within one main element. It demonstrates the present method, with or without adding the fifth-order reconstruction, can achieve positivity-preserving and oscillation-less solutions, and can fully exploit the subcell resolution of the DG method.

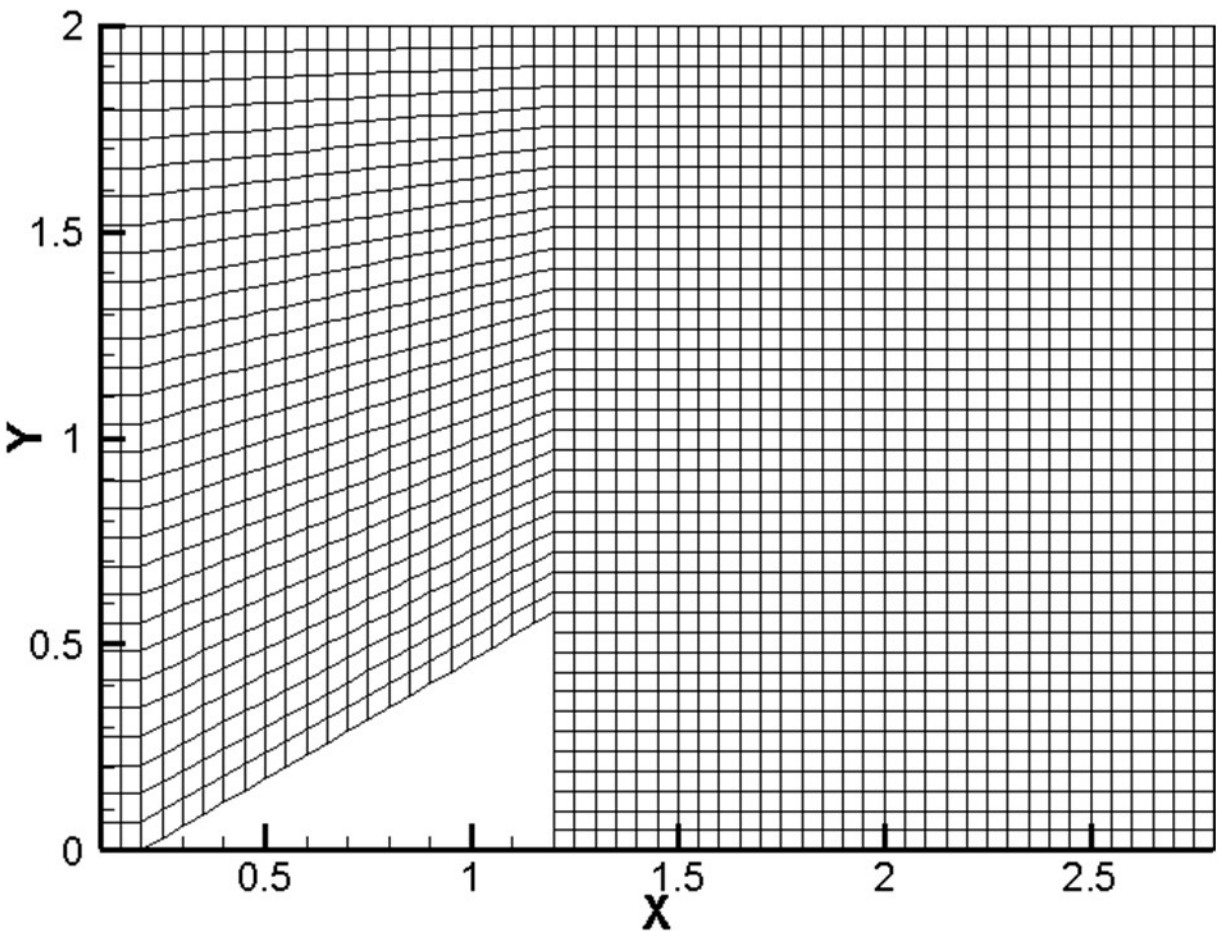

**Fig. 5** Mach 10 shock reflection and diffraction. Illustration of the domain and the rectangular mesh. In the figure the mesh size is around 1/20 that is 4 times the actual computation mesh size for better visual effects.

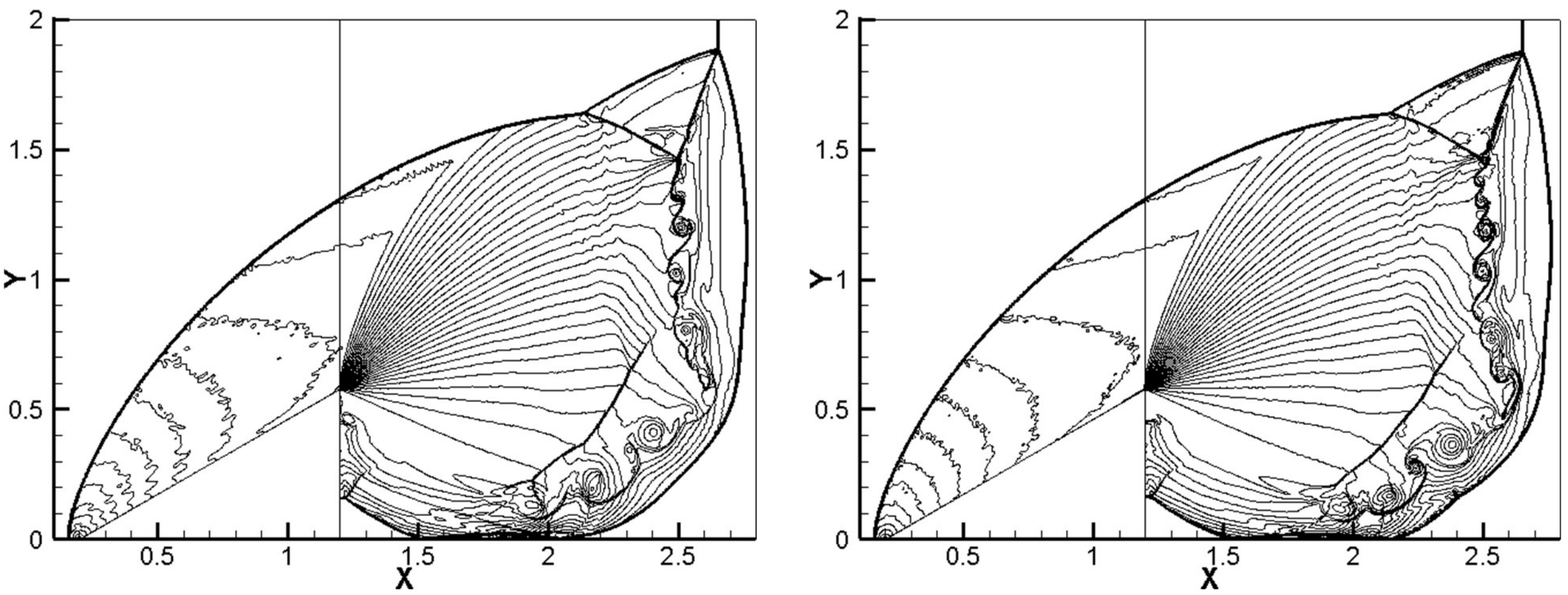

**Fig. 6** Mach 10 shock reflection and diffraction. Density solutions: 50 equally spaced contour lines from 0.05 to 25. Left: without the fifth-order reconstruction; right: with the fifth-order reconstruction.

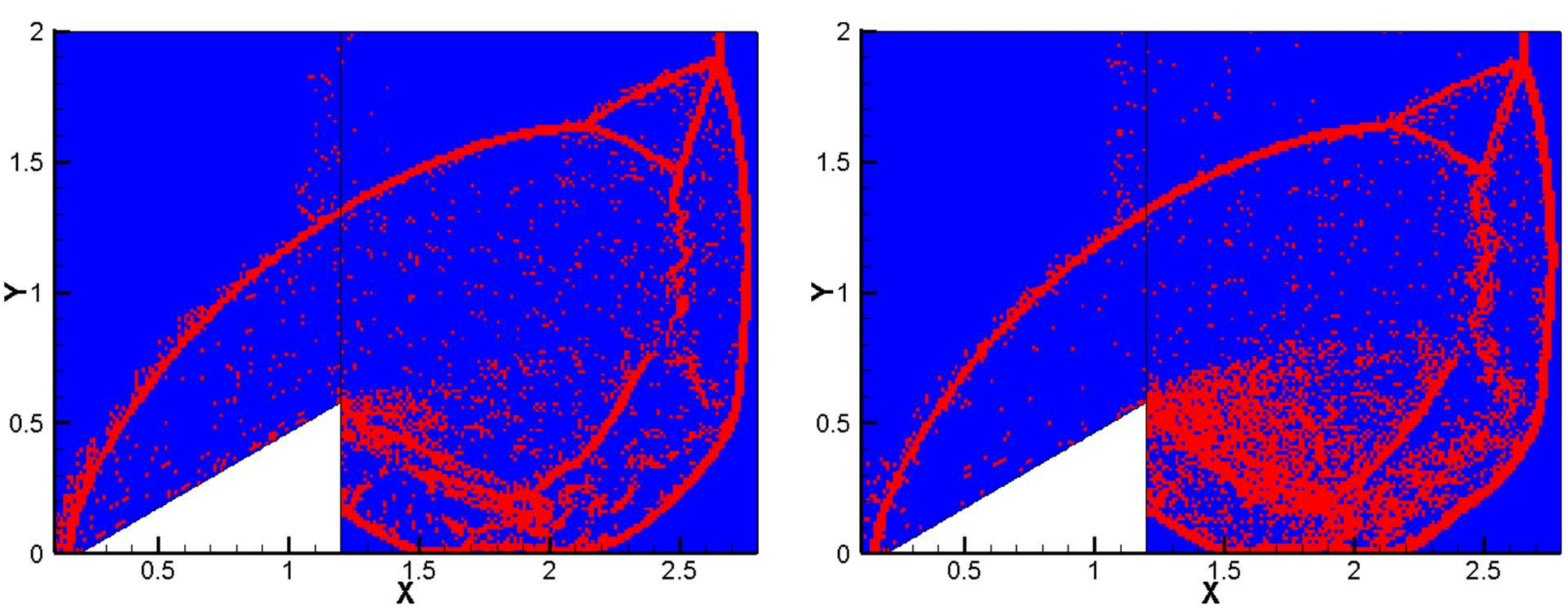

**Fig. 7** Mach 10 shock reflection and diffraction. The troubled cells are marked with red color. Left: without the fifth-order reconstruction; right: with the fifth-order reconstruction.

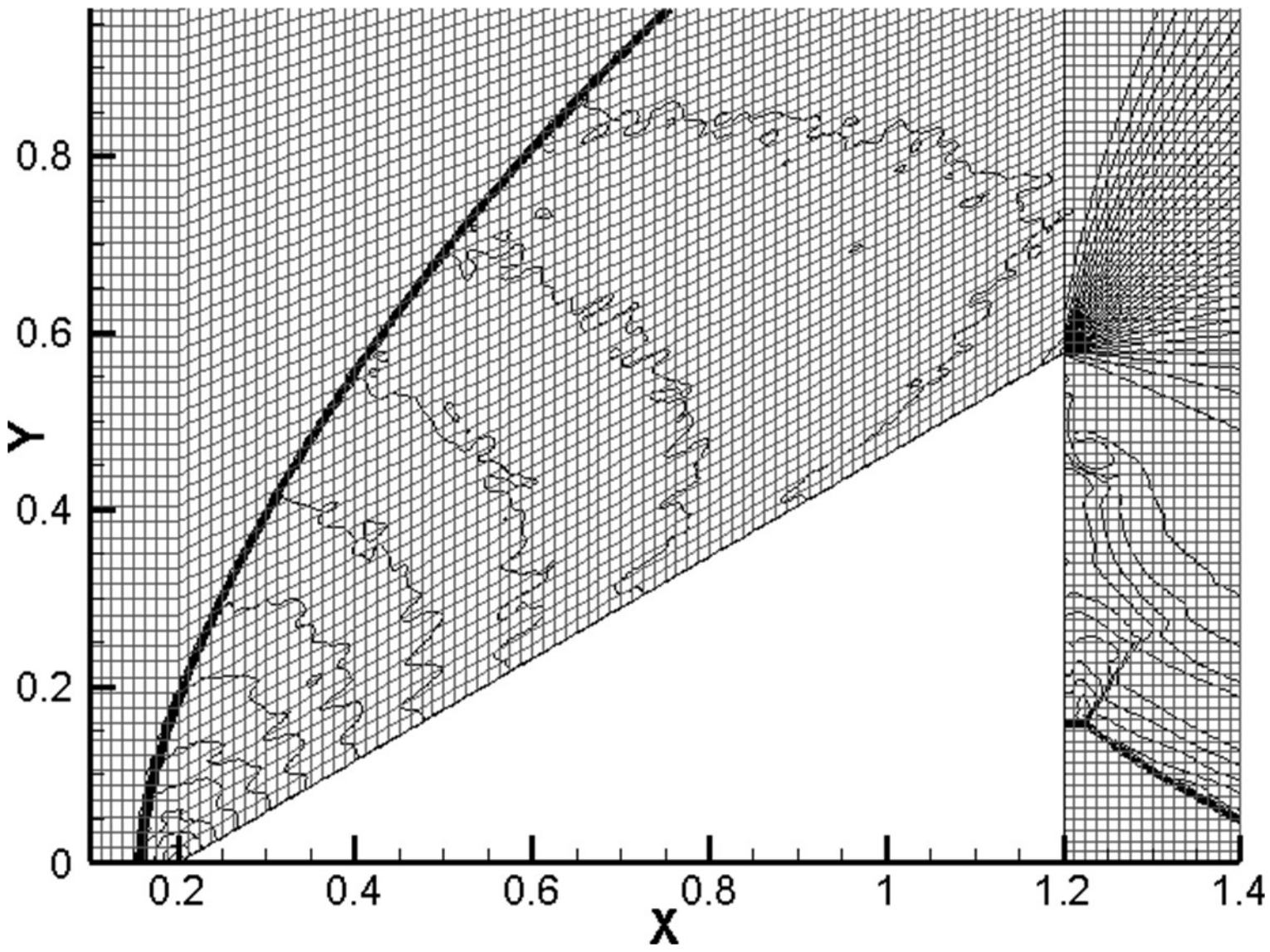

**Fig. 8** Mach 10 shock reflection and diffraction. Local view of the left figure of Fig. 6 with the computational grid.

### 4.2 Numerical results for the reactive Euler equations

In the following numerical experimentation, we apply the method to solve the reactive Euler equations. In some cases, the performance of the method is assessed by comparing it with the FV method using the same DOFs and the fifth-order WENO reconstruction scheme.

#### 4.2.1 One-dimensional detonation problems

Three one-dimensional detonation problems are considered in this subsection. The first example is the Chapman-Jouguet (C-J) detonation that has the chemical reaction modeled with an Arrhenius source term, see Appendix A. Similar cases are also considered in for instance [48, 49]. A computation domain of [0, 30] is full of burned and unburned gas, and initially, the burned gas with state ($\rho_{CJ}$, $u_{CJ}$, $p_{CJ}$, 0.0) is set on the left side of the computation domain, while the unburned gas with state ($\rho_0$, $u_0$, $p_0$, 1.0) is set on the right. The initial discontinuity is located at $x$=10. Given any initial state of $\rho_0$, $u_0$ and $p_0$, the C-J initial state of $\rho_{CJ}$, $u_{CJ}$ and $p_{CJ}$ can be determined according to [50]. Here we choose $\rho_0$=1.0, $u_0$=0.0 and $p_0$=1.0. The parameters are set as $\gamma$=1.4, $q_0$=25, $T_{ign}$=25 and $K_0$=16,418. A mesh with cell number $N$=100 is used for the DG/FV scheme and therefore the mesh with cell number $N$=400 is used for the WENO scheme. The computation evolves until final time $t$=1.8. The results of density, temperature and mass fraction are given in Fig. 9 where the reference solution is obtained using much refined meshes with 10000 cells. For the DG/FV scheme, the results are shown using the main element solution. The identified troubled cells are also marked in the first figure of Fig. 9 for the DG/FV scheme. The troubled cells are seen around the detonation wave. Comparing to the WENO scheme, the DG/FV scheme is able to capture the correct position of the wave.

The second example is the C-J detonation with the Heaviside model. The computation domain is [0, 0.05] and initially the discontinuity is set at $x$=0.005. We have the burned gas located at the region of $x$<0.005 and the unburned gas located at the region of $x$>0.005. The initial state for the unburned gas on the right is given by $\rho_0=1.201\times10^{-3}$, $u_0$=0.0, $p_0=8.321\times10^5$ and $z_0$=1.0. Then the C-J initial state for the burned gas on the left is again determined by [50]. The parameters are set as $\gamma$=1.4, $q_0=0.5196\times10^{10}$, $T_{ign}=0.1155\times10^{10}$ and $1/\xi=0.5825\times10^{10}$. Similar to the fist example, we utilize a mesh with cell number $N$=100 for the DG/FV scheme and $N$=400 for the WENO scheme. The computation evolves until final time $t=3\times10^{-7}$. The results of density, temperature and mass fraction are similarly given in Fig. 10. Again the identified troubled cells are only around the detonation wave and the DG/FV scheme has captured the correct position of the wave.

The third example taken from [51] involves a collision between a detonation wave and an oscillatory profile. The computation domain is [0, 2$\pi$] and the initial conditions are

$$\left(\rho,u,p,z\right)=\begin{cases}\left(1.79463,\ 3.0151,\ 21.53134,\ 0\right), & x<\pi/2,\\ \left(1+0.5\sin 2x,\ 0,\ 1,\ 1\right), & \mathit{otherwise}.\end{cases} \tag{22}$$

The parameters are $\gamma$=1.2, $q_0$=50, $T_{ign}$=3 and $1/\xi$=1000 in this case. We use a mesh with cell number $N$=80 for the DG/FV scheme and therefore $N$=320 for the WENO scheme. The computation is run for $t=\pi/5$ and the results of density, temperature and mass fraction are similarly plotted in Fig. 11. In this case the interaction of the detonation wave and oscillatory profile produces complicated flowfield. Nevertheless, the troubled cells are still identified only around the discontinuity. The smooth structures are well resolved while the location of the detonation front is correctly captured by the present method.

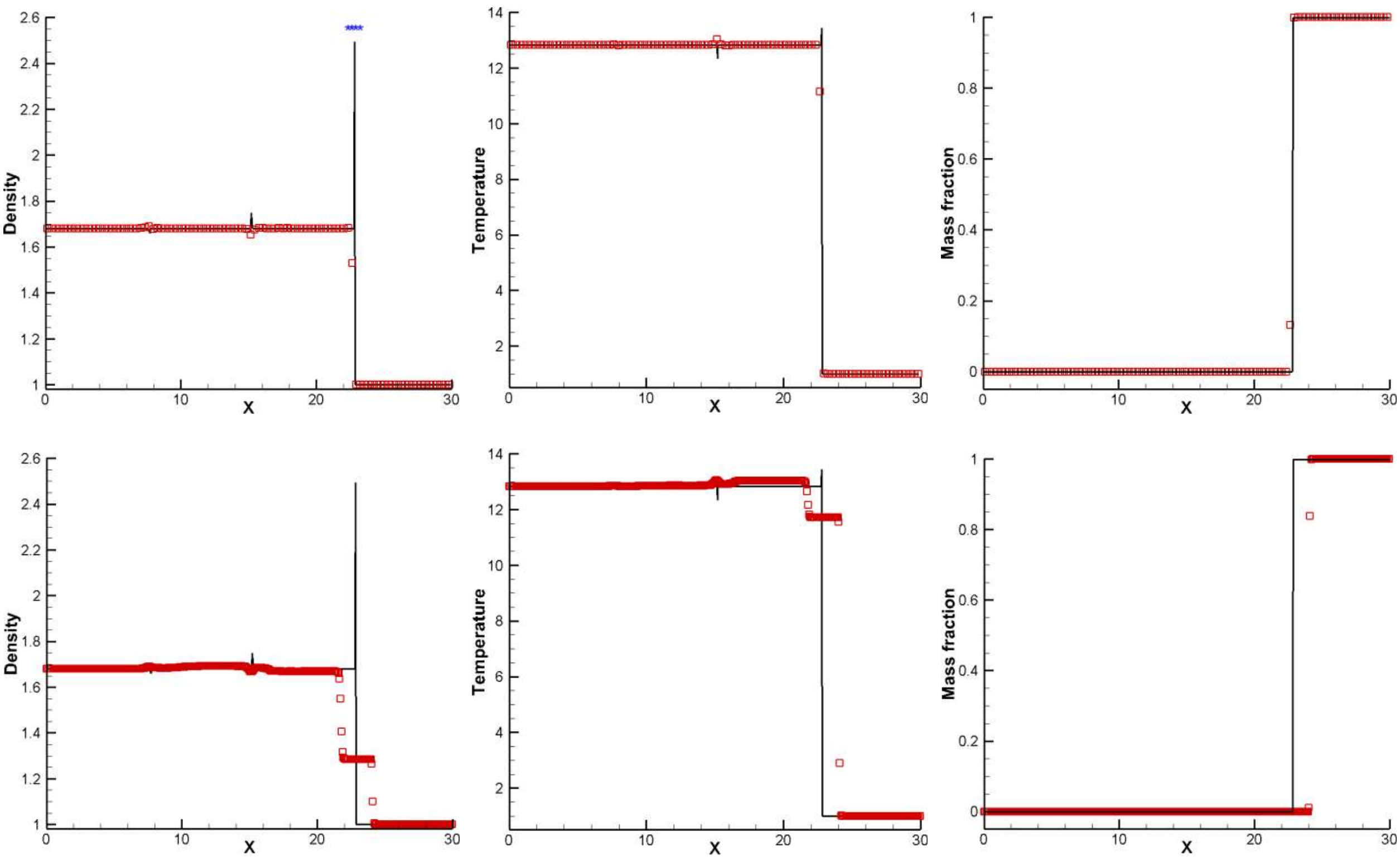

**Fig. 9** C-J detonation wave with the Arrehenius source. Results obtained by the DG/FV (upper) and WENO (lower) scheme. Bule asterisks in the first figure denote the identified troubled cells for the DG/FV scheme.

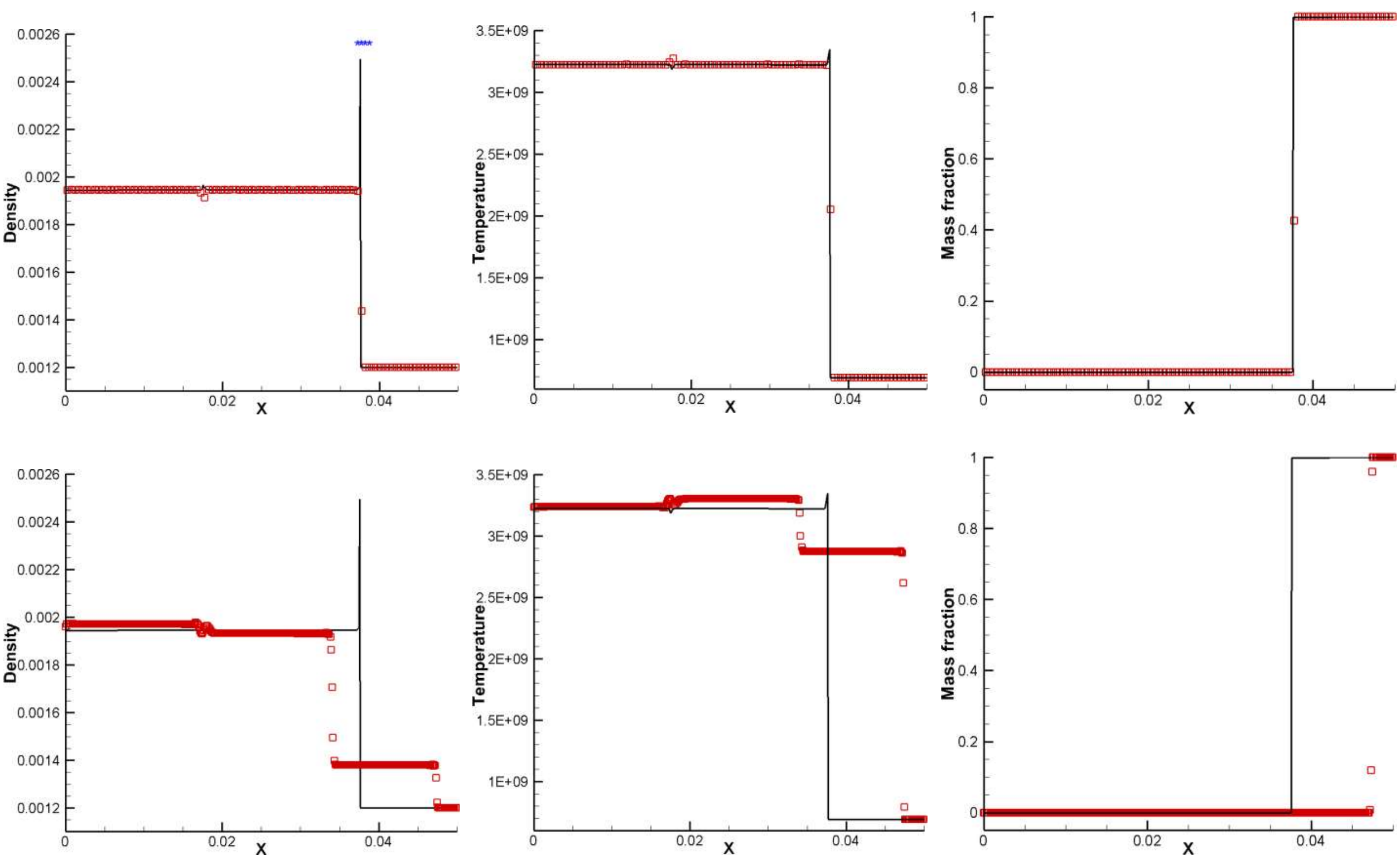

**Fig. 10** C-J detonation wave with the Heaviside source. Results obtained by the DG/FV (upper) and WENO (lower) scheme. Bule asterisks in the first figure denote the identified troubled cells for the DG/FV scheme.

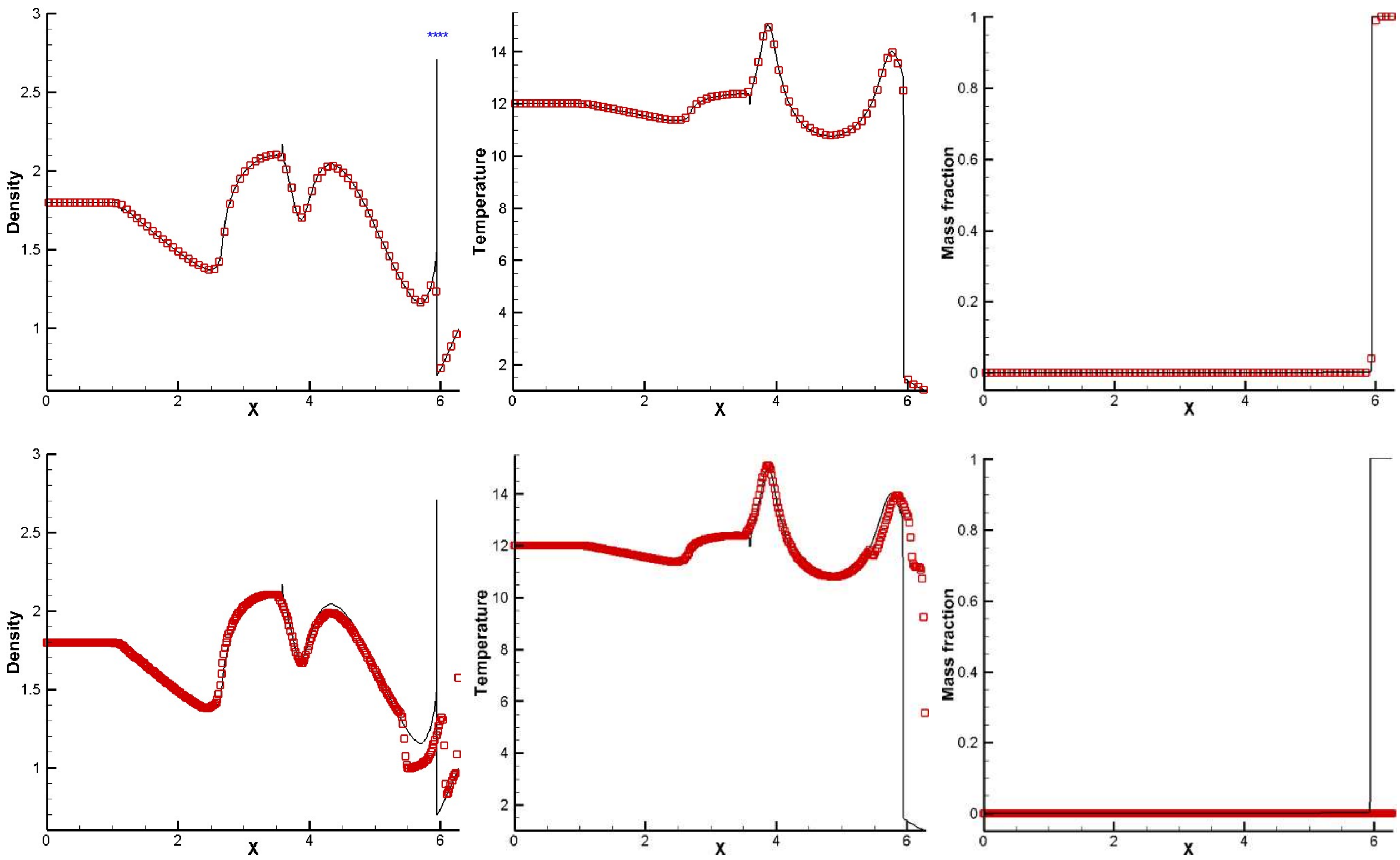

**Fig. 11** Collision of a detonation wave with an oscillatory profile. Results obtained by the DG/FV (upper) and WENO (lower) scheme. Bule asterisks in the first figure denote the identified troubled cells for the DG/FV scheme.

### 4.2.2 A 2D detonation wave

The 2D detonation wave problem, also investigated in for instance [1, 50, 52], is examined here. The computational domain is $(x, y)$=[0, 0.025]×[0, 0.005]. The initial conditions are

$$(\rho,u,v,p,\alpha)=\begin{cases}(\rho_l\ ,u_l,\ 0, p_l,\ 0), & if\ x\le\psi(y),\\ (\rho_r,u_r,\ 0,\ p_r,1), & otherwise,\end{cases} \tag{23}$$

where

$$\psi(y)=\begin{cases}0.004 & if\ \left|y-0.0025\right|\ge 0.001\\ 0.005-\left|y-0.0025\right| & if\ \left|y-0.0025\right|<0.001\end{cases} \tag{24}$$

The Heaviside form with same parameters $\gamma$, $q_0$, $T_{\text{ign}}$ and $1/\xi$ as the second example in Sec. 4.2.1 is used to model the chemical reaction. The right states are $\rho_r$=1.201×10$^{-3}$, $u_r$=0.0, $p_r$=8.321×10$^{5}$ which are also the same as the second example in Sec. 4.2.1. The same states can be determined by the C-J detonation model for $\rho_l=\rho_{CJ}$, $p_l=p_{CJ}$, whereas $u_l$=8.162×10$^{4}$>$u_{CJ}$. One important feature of this case is the appearance of triple points that travel in the transverse direction and reflect back and forth from upper and lower boundaries. We carry out the numerical experiment using 200×40 elements for the DG/FV scheme, therefore 800×160 elements for the WENO scheme. A solution from [25] using 2000×400 elements is adopted here as the reference solution. The density contours with the subcell data for DG/FV are compared with WENO at the evolutionary time $t$=0.3×10$^{-7}$, $t$=0.92×10$^{-7}$ and $t$=1.7×10$^{-7}$ in Figs. 12-14. We also give the density distribution along $y$=0.0025 at $t$=1.7×10$^{-7}$ in Fig. 15. The density solutions of the DG/FV method are without spurious oscillations at the subgrid level. The method is able to achieve comparable and even better solutions compared to the WENO scheme. For example, it can be seen that fewer spurious waves are produced in front of the detonation in Fig. 13 and better agreement with the reference solution is obtained in Fig. 15.

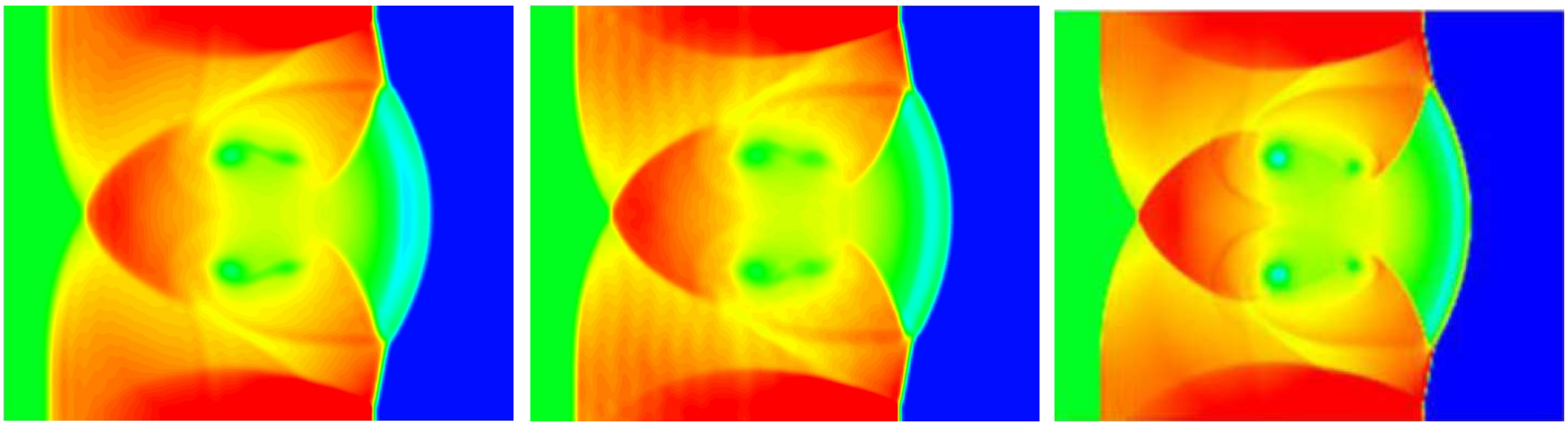

**Fig. 12** 2D detonation wave. Density counters at $t=0.3\times10^{-7}$ for WENO (left), DG/FV (middle) and reference solution (right).

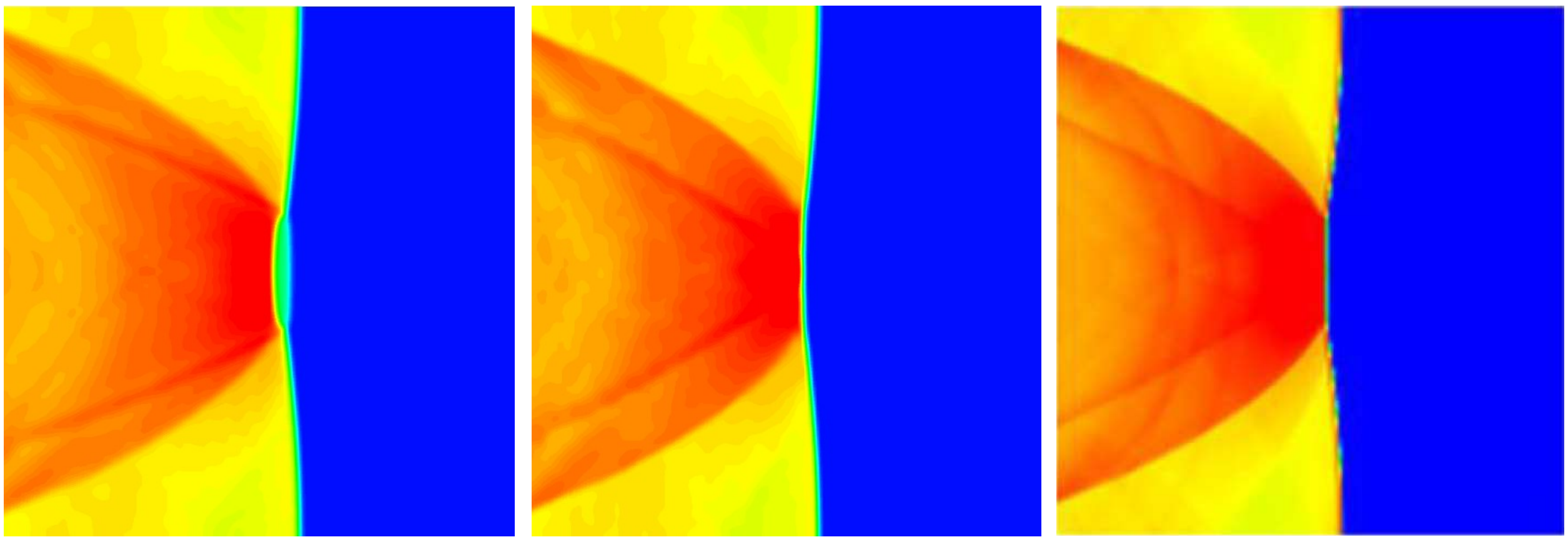

**Fig. 13** 2D detonation wave. Density counters at $t=0.92\times10^{-7}$ for WENO (left), DG/FV (middle) and reference solution (right).

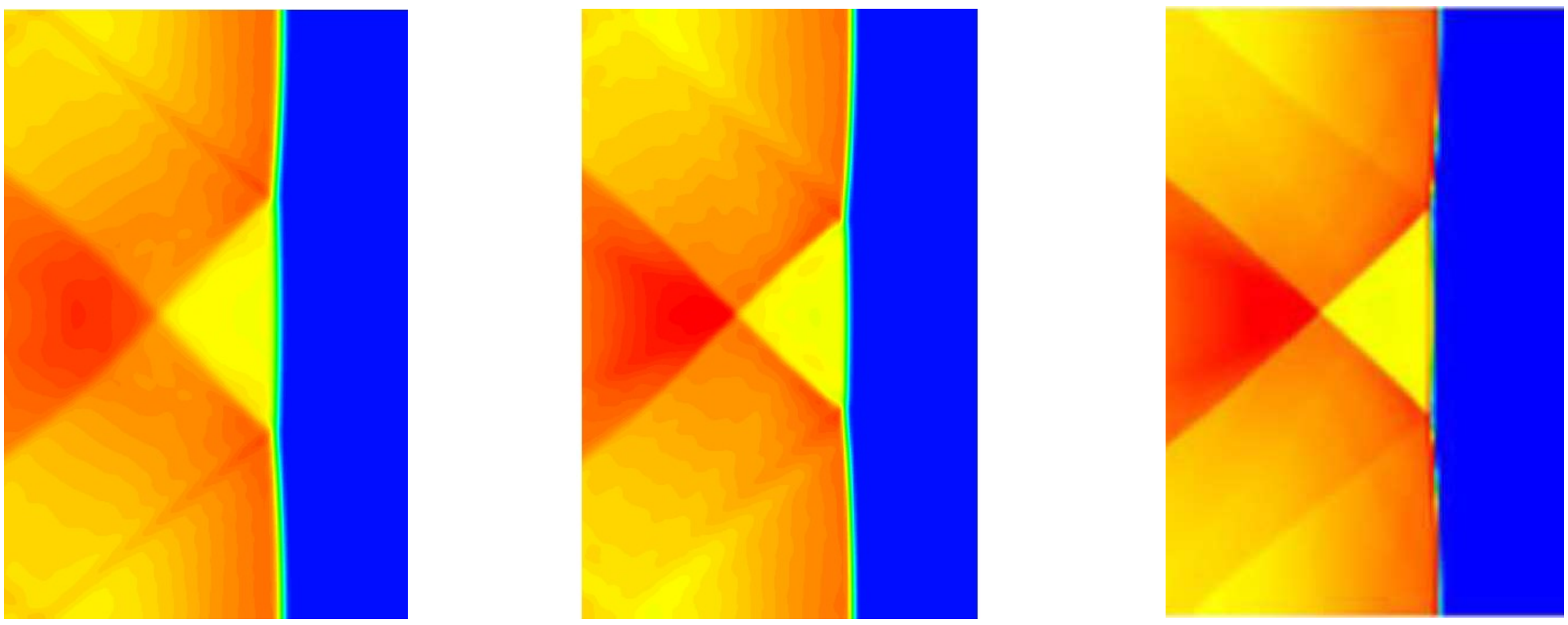

**Fig. 14** 2D detonation wave. Density counters at $t=1.7\times10^{-7}$ for WENO (left), DG/FV (middle) and reference solution (right).

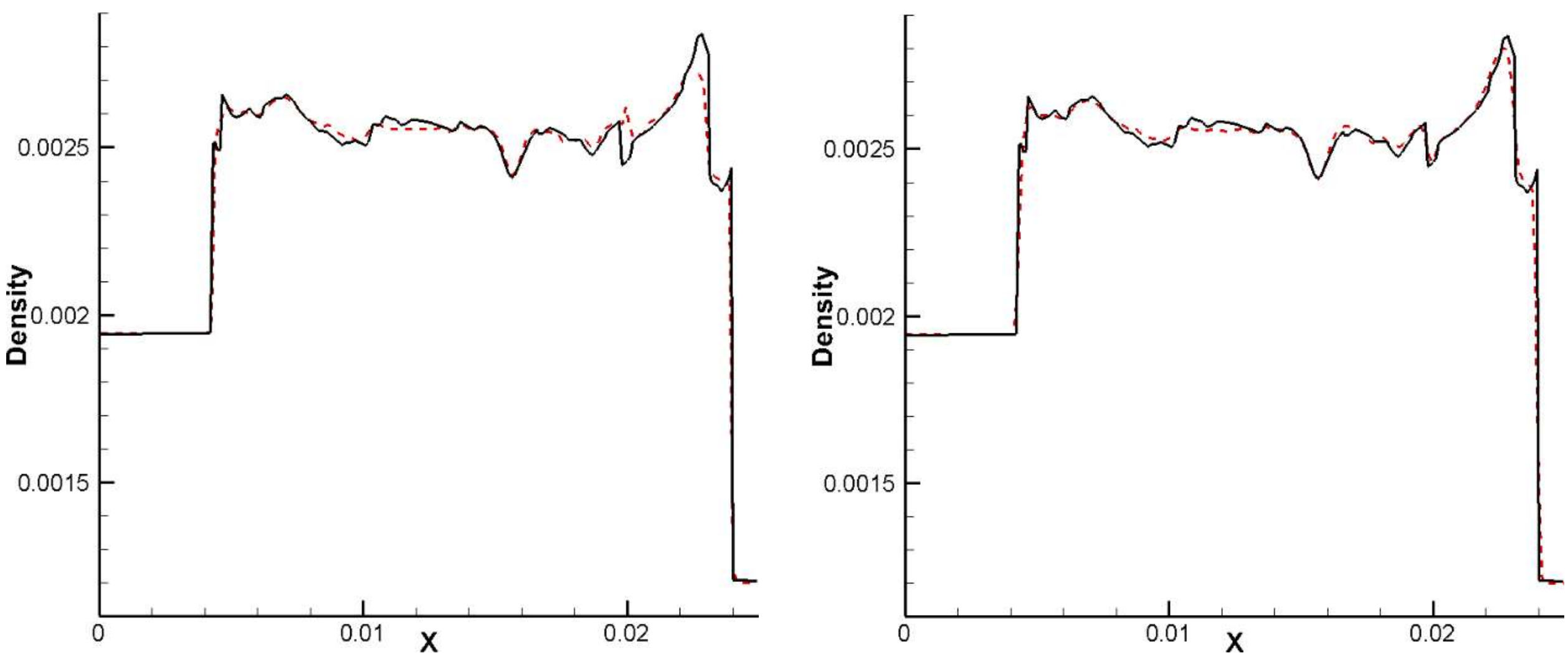

**Fig. 15** 2D detonation wave. 1D cross-section along the central line obtained by the WENO (left) and DG/FV (right) method. Solid line: reference solution; dashed line: present computation with the WENO (left) or DG/FV (right) method.

### 4.2.3 Detonation diffraction problem with 90° corner

The detonation wave passing through an obstacle with an angle of ninety degrees, see also [21, 53]. The physical domain is $(x, y)=[0, 5]\times[0, 5]$, which includes the obstacle area of $(x, y)=[0, 1]\times[0, 2]$. The initial conditions are

$$\left(\rho, u, v, \rho e, z\right)=\begin{cases}\left(11,\ 6.18,\ 0, 970,\ 1\right), & x<0.5,\\ \left(1,\ 0,\ 0,\ 55,\ 1\right), & otherwise.\end{cases} \tag{25}$$

The parameters are given as $\gamma$=1.2, $q_0$=50, $T_{\text{ign}}$=50 and $K_0$=2566.4. The reflective boundary conditions are applied at every boundary except that at $x$=0, where we have $(\rho, u, v, \rho e, z)$=(11, 6.18, 0, 970, 1). The numerical experiment is carried out on a uniform mesh with $\Delta x=\Delta y$=1/48 for the DG/FV scheme and $\Delta x=\Delta y$=1/192 for the WENO scheme. In this case, the WENO scheme has to be equipped with the positivity-preserving strategy in order to handle the sudden drop of the pressure or density close to zero. The strategy to preserve the positivity for the WENO scheme is to use a similar *a posteriori* computation process described in Sec. 3.3. Once the negative values are detected in the reconstructed interface variables or the computed candidate solutions, the first-order Godunov scheme is performed until no negative values are detected for the candidate solutions. The computation lasts until time $t$=0.6 and the contours of density and pressure are displayed in Figs. 16-17. The solutions on the subcell obtained by the DG/FV method are less oscillatory. Small oscillations ahead of the second shock are seen for the WENO method, which are likely caused by the current implementation without the characteristic decomposition. A very close look actually reveals slightly sharper shock transition obtained by the DG/FV method. Nevertheless in general, the two methods gain comparable results in this test case. In Fig. 18 the identified troubled cells are on the whole around discontinuities, indicating a successful implementation of the troubled cell indicator.

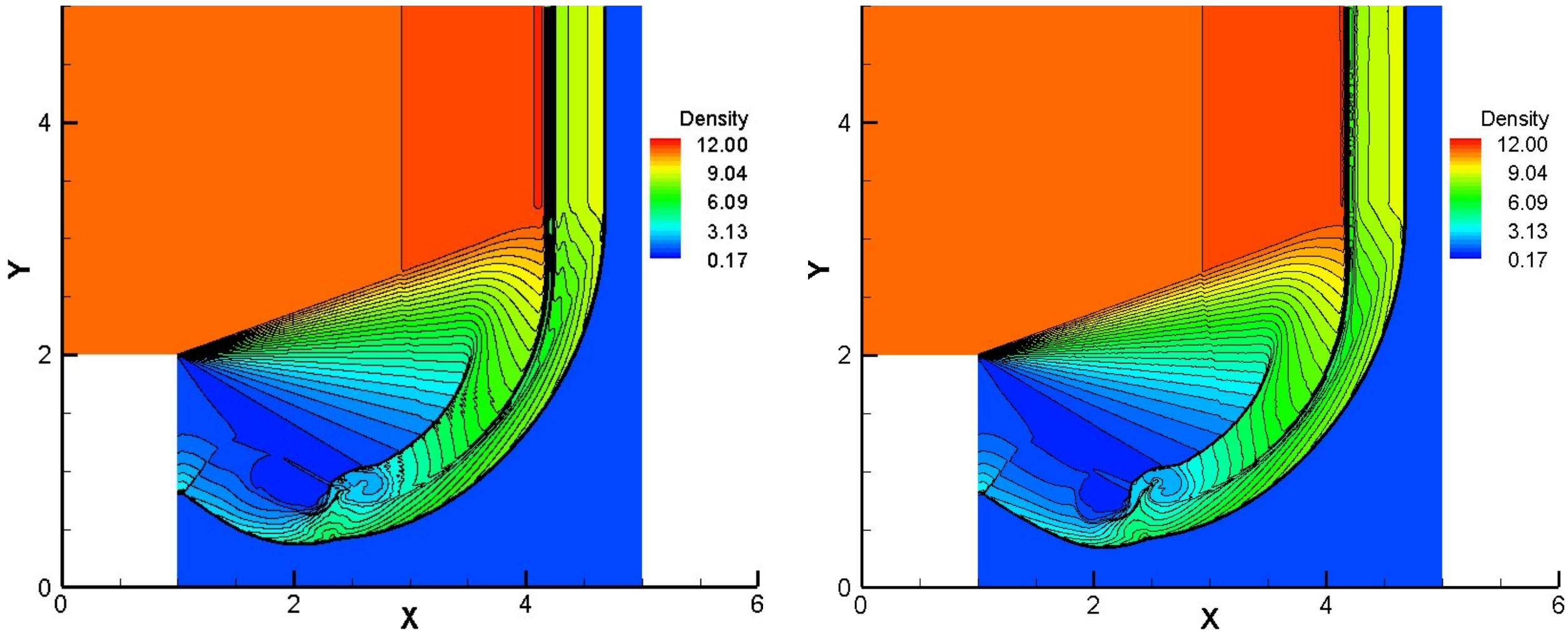

**Fig. 16** Detonation diffraction problem with 90° corner. Density counters.
Left: solutions obtained by the WENO method; right: solutions obtained by the DG/FV method.

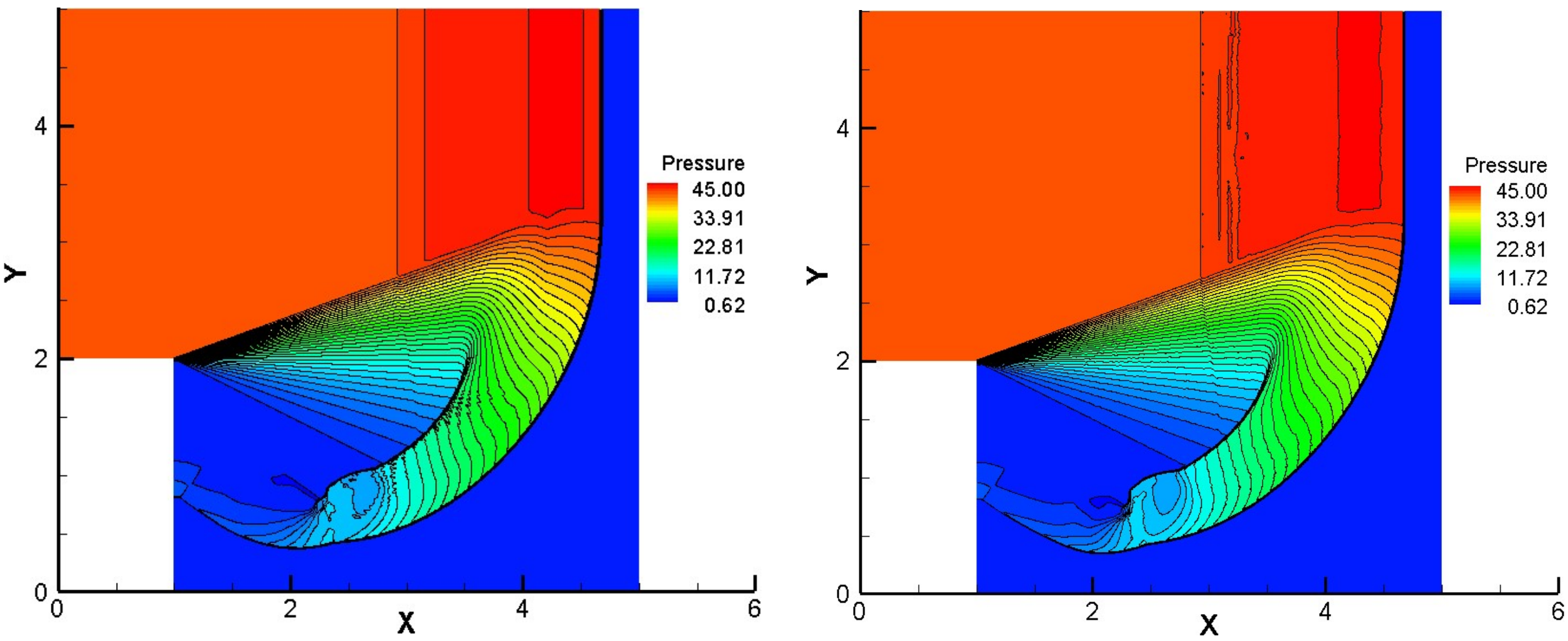

**Fig. 17** Detonation diffraction problem with 90° corner. Pressure counters.
Left: solutions obtained by the WENO method; right: solutions obtained by the DG/FV method.

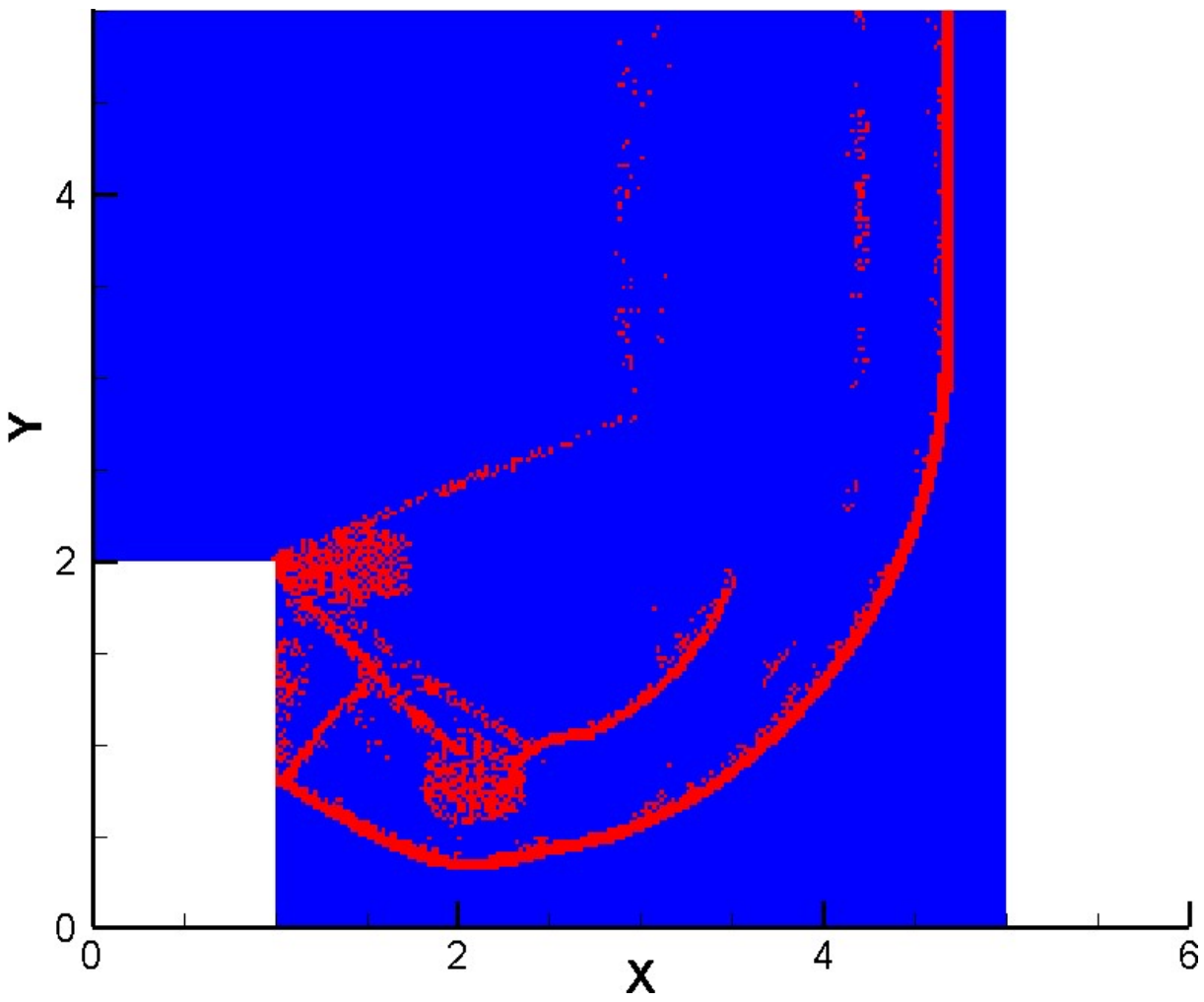

**Fig. 18** Detonation diffraction problem with 90° corner. Identified troubled cells for the DG/FV method are marked with red color.

#### 4.2.4 Multiple obstacles

The example is the detonation wave passing multiple rectangular obstacles, see for instance [21, 53]. The physical domain is $(x, y)$=[0, 8.3]×[0, 10]. The first obstacle is located at [1.3, 3.3]×[0, 2.6] and the second one is located at [5.1, 8.3]×[0, 4.3]. The initial conditions are

$$(\rho, u, v, \rho e, z) = \begin{cases} (7,\ 0,\ 0,\ 200,\ 0), & x^2 + y^2 \le 0.36, \\ (1,\ 0,\ 0,\ 55,\ 1), & \textit{otherwise.} \end{cases} \tag{26}$$

The parameters are given as $\gamma$=1.2, $q_0$=50, $T_{ign}$=20 and $K_0$=2410.2. The reflective boundary conditions are applied everywhere. The computational mesh of the DG/FV method is uniformly rectangular with $\Delta x$=$\Delta y$=1/20 that is again coarser than most meshes used for this test case. Fig. 19 displays the contours of pressure and density in the subgrid level at final time $t$=1.4. In Fig. 19 one can see that the multiple shock waves are captured without obvious oscillations. Some interesting features of the density solution are observed at the left side of the first obstacle, which is overall comparable to those shown in Fig. 4.9(b) in [21]. The DG/FV scheme seems to obtain results close to those published in [21, 53]. The identified troubled cells are also on the whole around various discontinuities in Fig. 20.

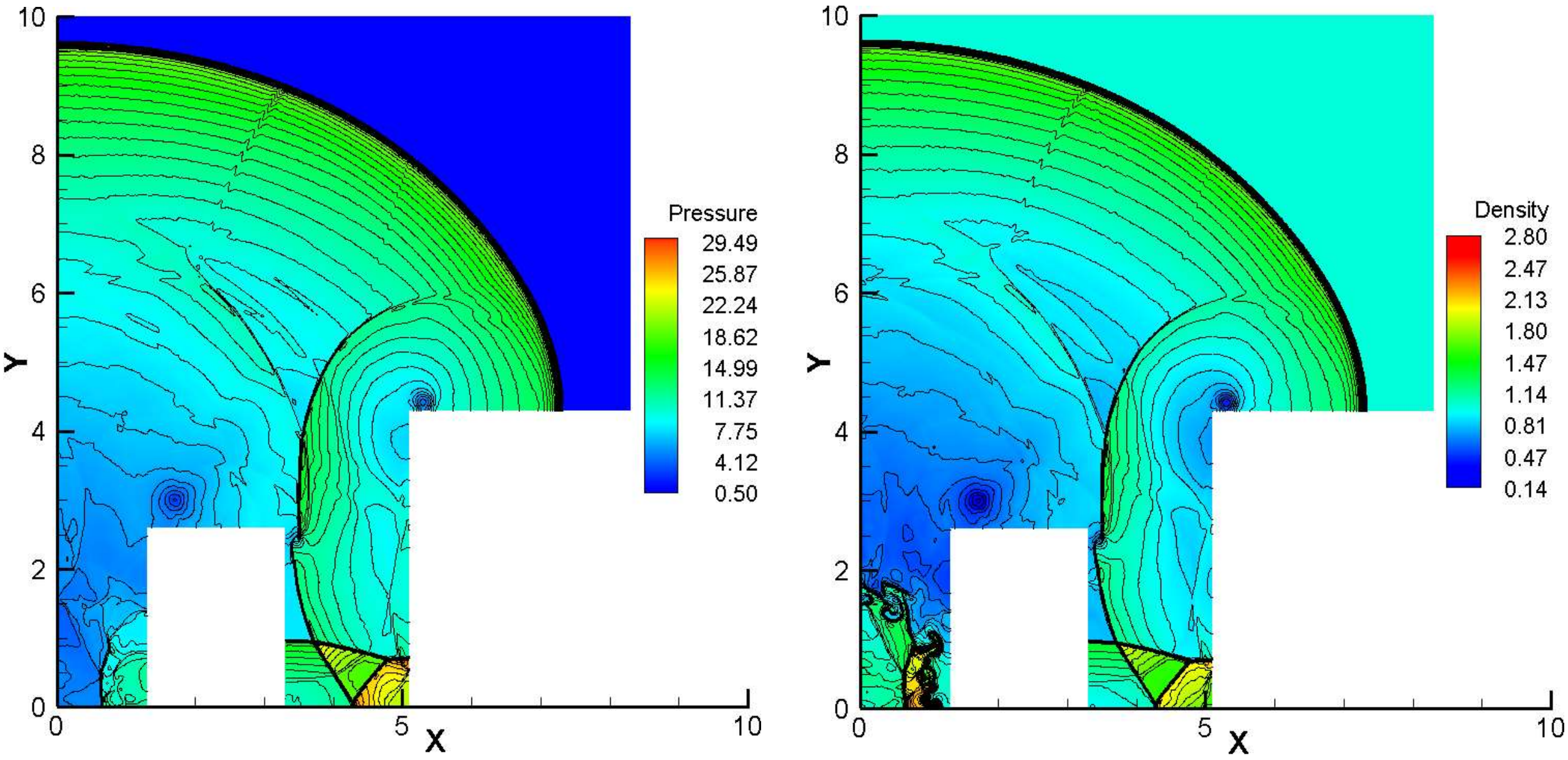

**Fig. 19** Multiple obstacles problem. Solutions obtained by the DG/FV method. Left: pressure counters; right: density counters.

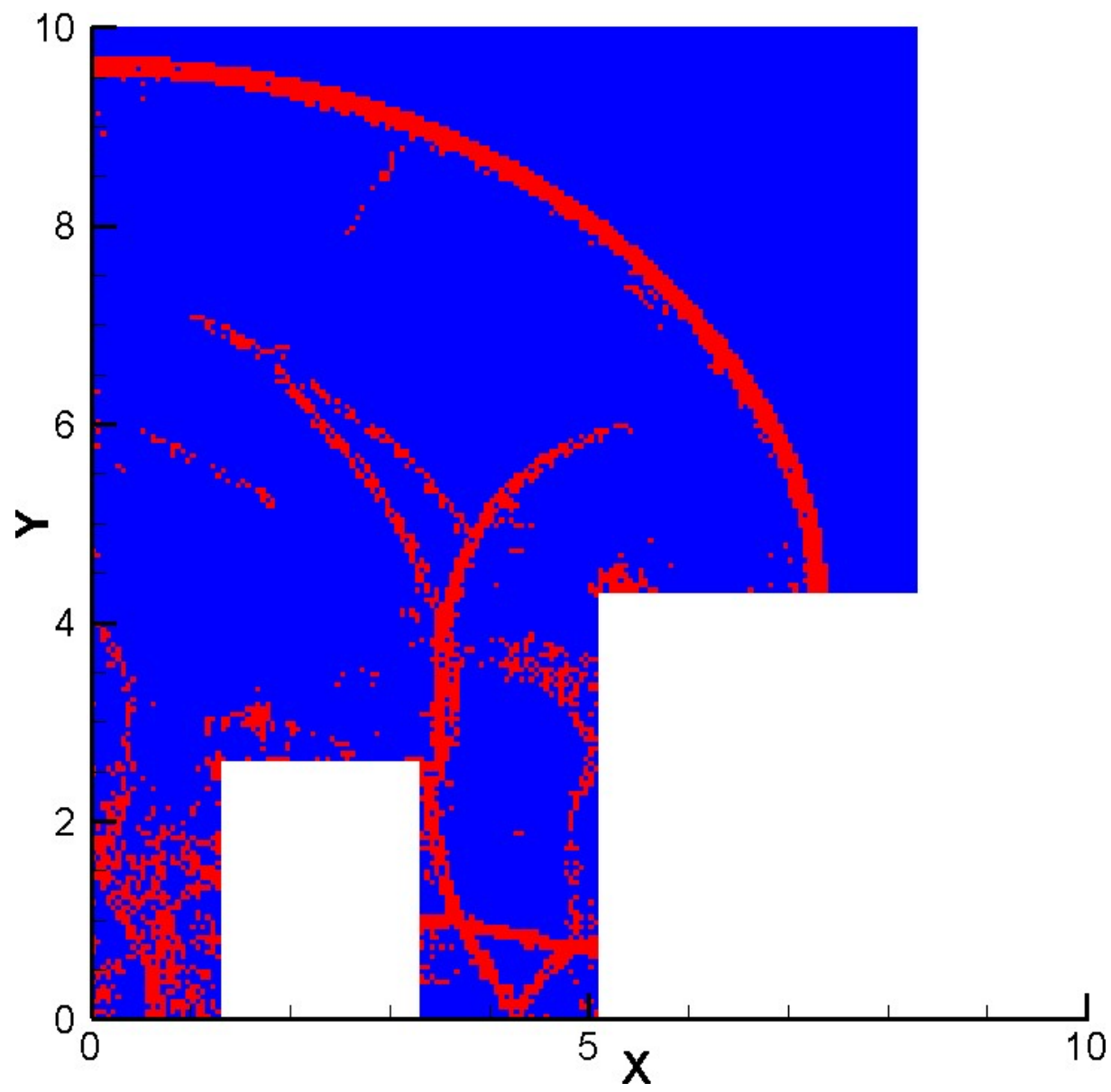

**Fig. 20** Multiple obstacles problem. Identified troubled cells for the DG/FV method are marked with red color.

### 4.2.5 Shock focusing

This test case, see also [54-56], is to test the performance of the method in using irregular quadrilateral meshes to solve the gaseous detonation problem. The computational domain shown in Fig. 21 is an irregular shaped region composed of three straight edges and a curve edge of a parabola of $y^2=-2.7x$. The reflective boundary condition is applied everywhere except that at $x=-5$ the supersonic inflow boundary condition is adopted. There are 115 points equally distributed between (−5, 2.7) and (−2.7, 2.7), 135 points equally distributed between (−2.7, 2.7) and (−1, $\sqrt{2.7}$), 135 points equally distributed between (−1, $\sqrt{2.7}$) and (0, 0), 250 points equally distributed between (0, 0) and (−5, 0), and 135 points equally distributed between (−5, 0) and (−5, 2.7). Therefore the computational meshes have a total of 250×135 quadrilateral elements. The grids have to be deformed in order to fill the whole region, see Fig. 21 for the details of local irregular grids around the connection point (−1, $\sqrt{2.7}$). At the initial time, a shock wave with $Ma$=2.5 moving to the right is located at $x=-2.7$. The preshock state is $(\rho, u, v, p, z)$=(0.2, 0.0, 0.0, 0.2, 1.0) and the source term is in an Arrhenius form with $\gamma$=1.4, $q_0$=50, $T_{ign}$=20 and $K_0$=30. Similar parameters are also considered in for instance [56]. From the temperature results at $t$=1.8 in Fig. 22, the detonation waves and complex combustion structures appear in the flow field, see for instance Fig. 3.8(f) in [56] for a comparison. Despite the use of a coarse mesh with cell length $h\approx$0.015-0.02, both discontinuous and smooth flow structures are well captured. Fig. 22 also displays the results obtained by the DG/FV method without using the adaptive reconstruction, i.e. only using the THINC reconstruction with $\beta$=1.1 for the FV computation on the troubled cells. By comparing results in Fig. 22 to those in Fig. 3.8(e)-(f) in [56], it seems to us that the method using the THINC reconstruction with $\beta$=1.1 achieves similar resolution to the results on the coarse mesh in [56], while the method using the adaptive reconstruction achieves similar resolution to the results on the refined mesh in [56]. It highlights the effectiveness of the present method by hybridizing the THINC scheme with anti-diffusion effect ($\beta$=1.6). In Fig. 23, the local enlargement of the left figure of Fig. 22 indicates that the discontinuities have been captured within the width of one main element, which is also the proof of exploiting the multiscale resolution of the DG method in the detonation simulation. In Fig. 24 the identified troubled cells are seen around various discontinuities and some complex flow structures, which do not affect the ability of the method to capture delicate flow details.

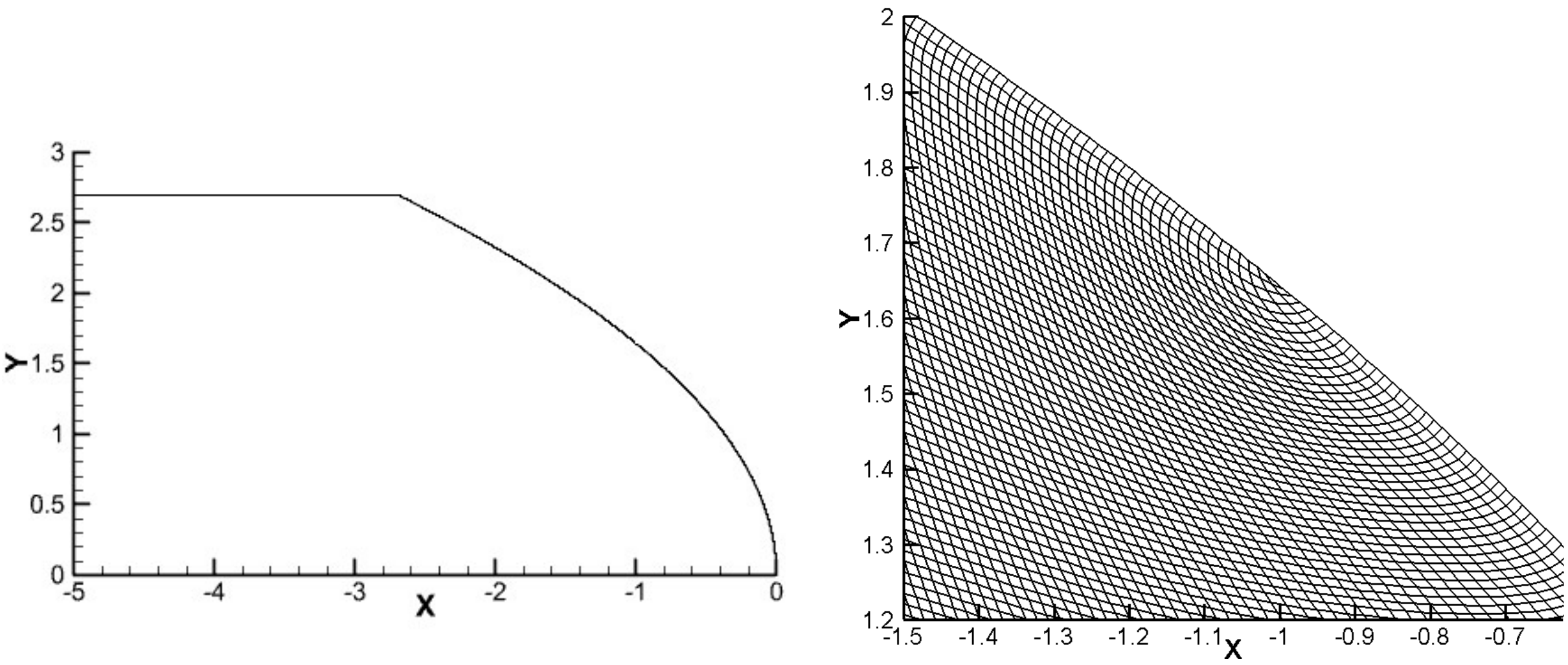

**Fig. 21** Shock focusing problem. The computational domain and the local view of the computational meshes.

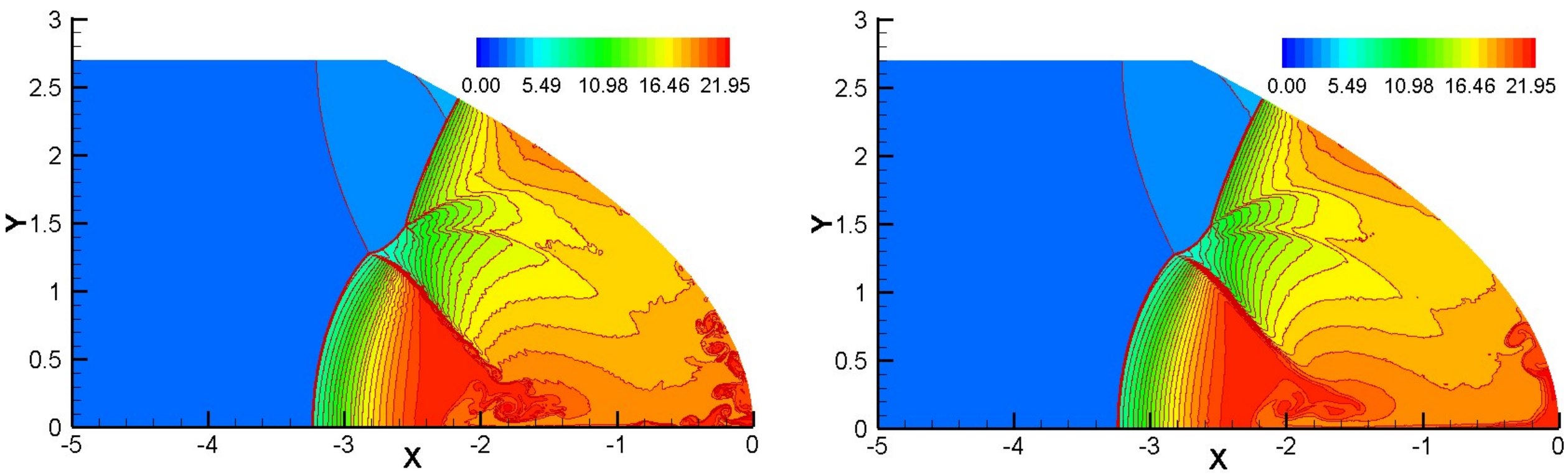

**Fig. 22** Shock focusing problem. Temperature solutions with 29 contour lines. Left: results by the present method using the adaptive reconstruction on the troubled cells; right: results by the method using the THINC reconstruction with $\beta$=1.1 on the troubled cells.

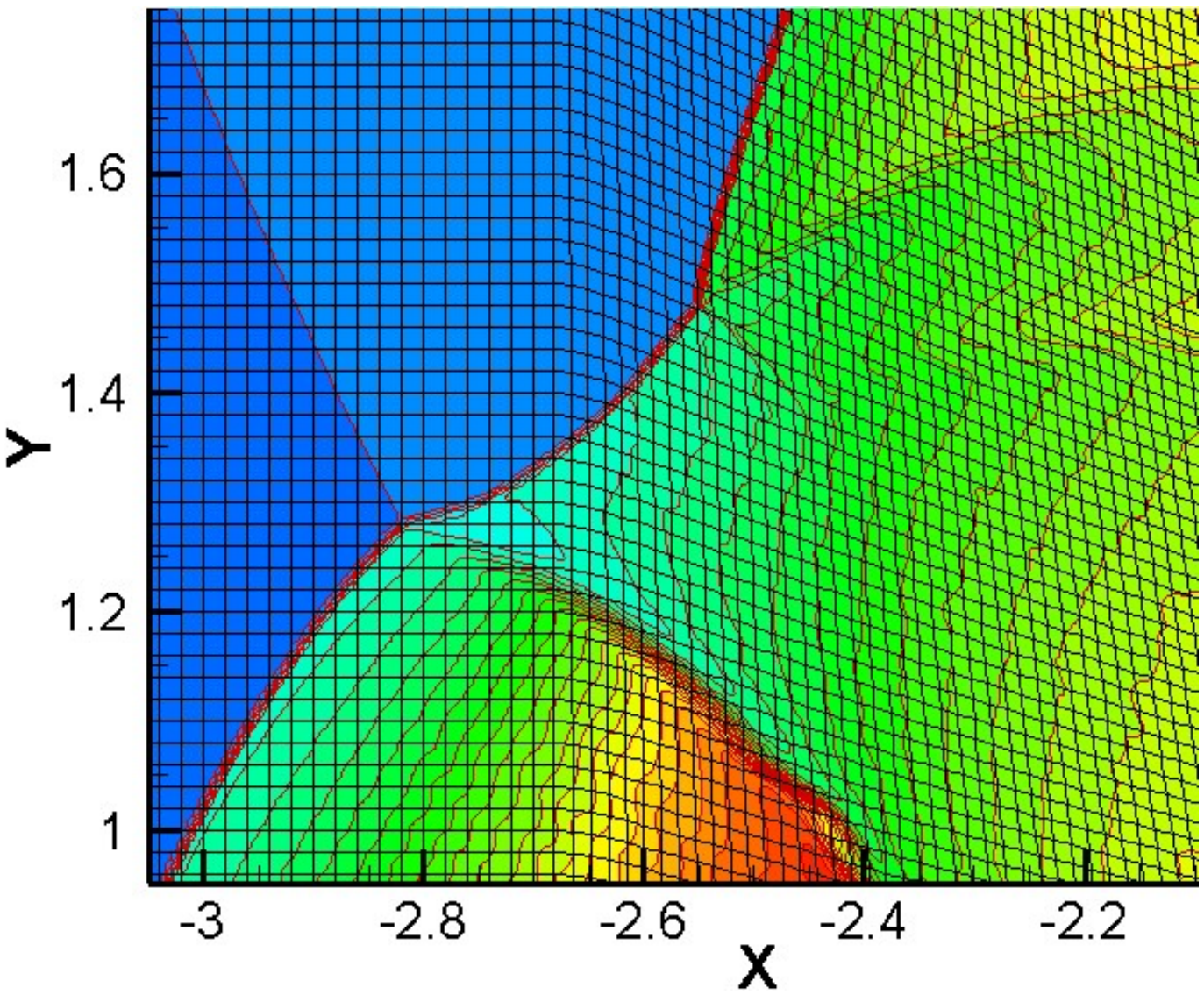

**Fig. 23** Shock focusing problem. Local view of the left figure of Fig. 22 with the computational grid.

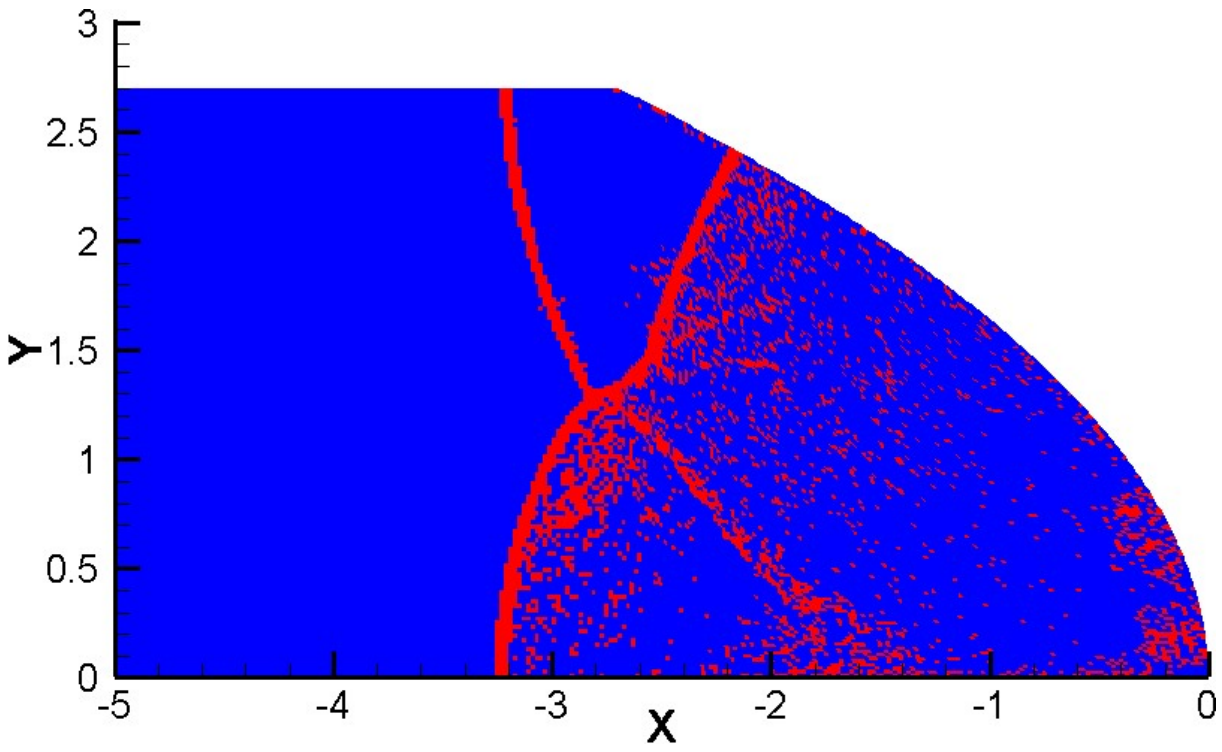

**Fig. 24** Shock focusing problem. Identified troubled cells for the present method are marked with red color.

### 4.2.6 A multi-species single-reaction radially symmetric detonation wave

This is a 2D radially symmetric detonation wave with four species and one reaction. The reaction model for multi-species and multi-reactions can be found in Appendix B. In this test case the prototype reacting model is

$$\mathrm{CH_4} + 2\mathrm{O_2} \rightarrow \mathrm{CO_2} + 2\mathrm{H_2O} \tag{27}$$

Similar test cases are also considered in for example [57]. The parameters are given as $\gamma$=1.4, $T_1$=2, $B_1$=$10^6$, $q_1$=200, $q_2$=$q_3$=$q_4$=0, $M_1$=16, $M_2$=32, $M_3$=44 and $M_4$=18. The computational domain is ($x$, $y$)=[0, 50]×[0, 50]. Initially, totally burnt gas exists in a circle of radius 10, while totally unburnt gas is everywhere outside of the circle. The initial condition is thus set as

$$(\rho, u, v, p, z_{\mathrm{CH_4}}, z_{\mathrm{O_2}}, z_{\mathrm{CO_2}}, z_{\mathrm{H_2O}}) = \begin{cases} (2,\ 10x/r,\ 10y/r,\ 40,\ 0,\ 0.2,\ 0.475,\ 0.325), & r \le 10, \\ (1,\ 0,\ 0,\ 1,\ 0.1,\ 0.6,\ 0.2,\ 0.1), & r > 10, \end{cases} \tag{28}$$

where $r = \sqrt{x^2 + y^2}$. The solid-wall boundary condition is applied at $x$=0 and $y$=0. The outflow boundary condition is applied at $x$=50 and $y$=50. The fourth-order DG/FV is carried out on a 25×25 structured mesh that is intentionally deformed, see Fig. 25, while the FV using the fifth-order WENO is conducted on the subcells of the same 25×25 mesh, resulting in 100×100 DOFs for both methods. The computation is stopped until time $t$=6.0. We give the velocity fields at $t$=2.0, $t$=4.0 and $t$=6.0 in Fig. 26, where the detonation front is supposed to be circular. The present method has captured the moving circular detonation front nicely, while the spurious wave appears for the WENO scheme. The density results at the cross-section $y$=$x$ at $t$=4 are compared in Fig. 27 where the exact solution is obtained by the fifth-order WENO (WENO5) with a much finer mesh of 1000×1000. The detonations are well-captured by the DG/FV method on the coarse mesh compared to the WENO scheme under the same DOFs.

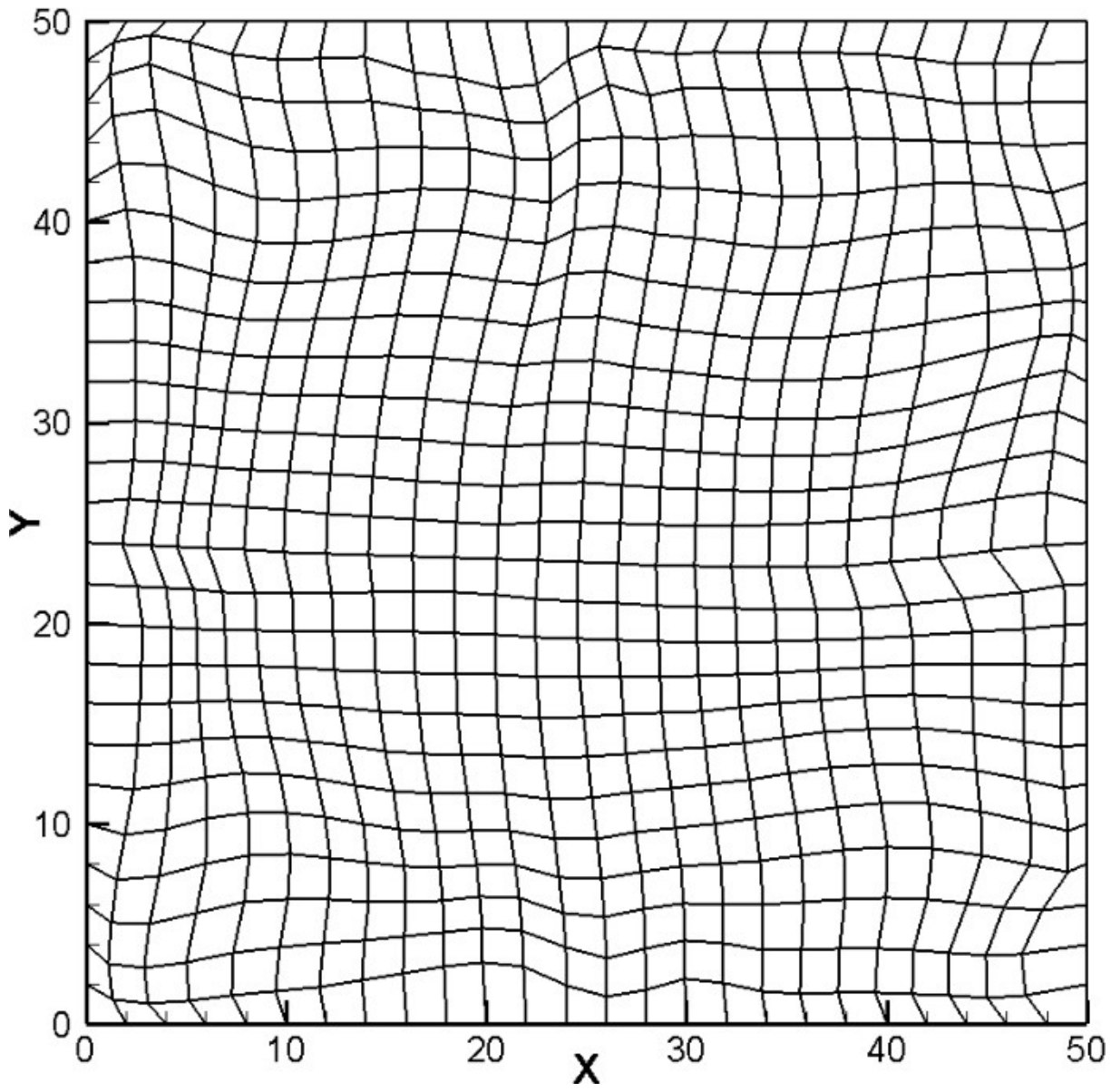

**Fig. 25** Radially symmetric detonation wave problem. The computational domain and computational mesh.

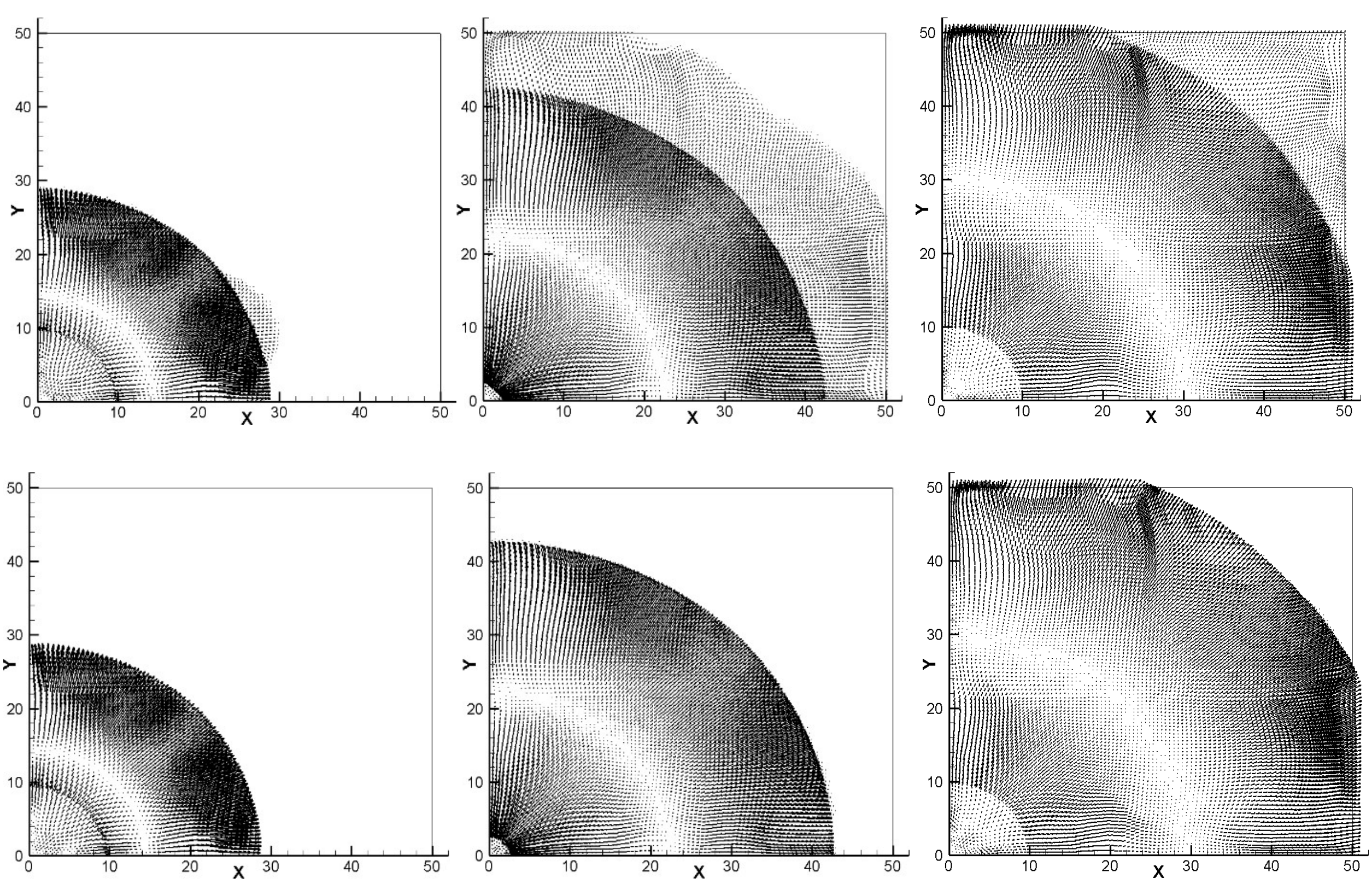

**Fig. 26** Radially symmetric detonation wave problem. Velocity fields at $t$=2, $t$=4 and $t$=6 obtained by the WENO (upper) and DG/FV (lower) method using the same DOFs.

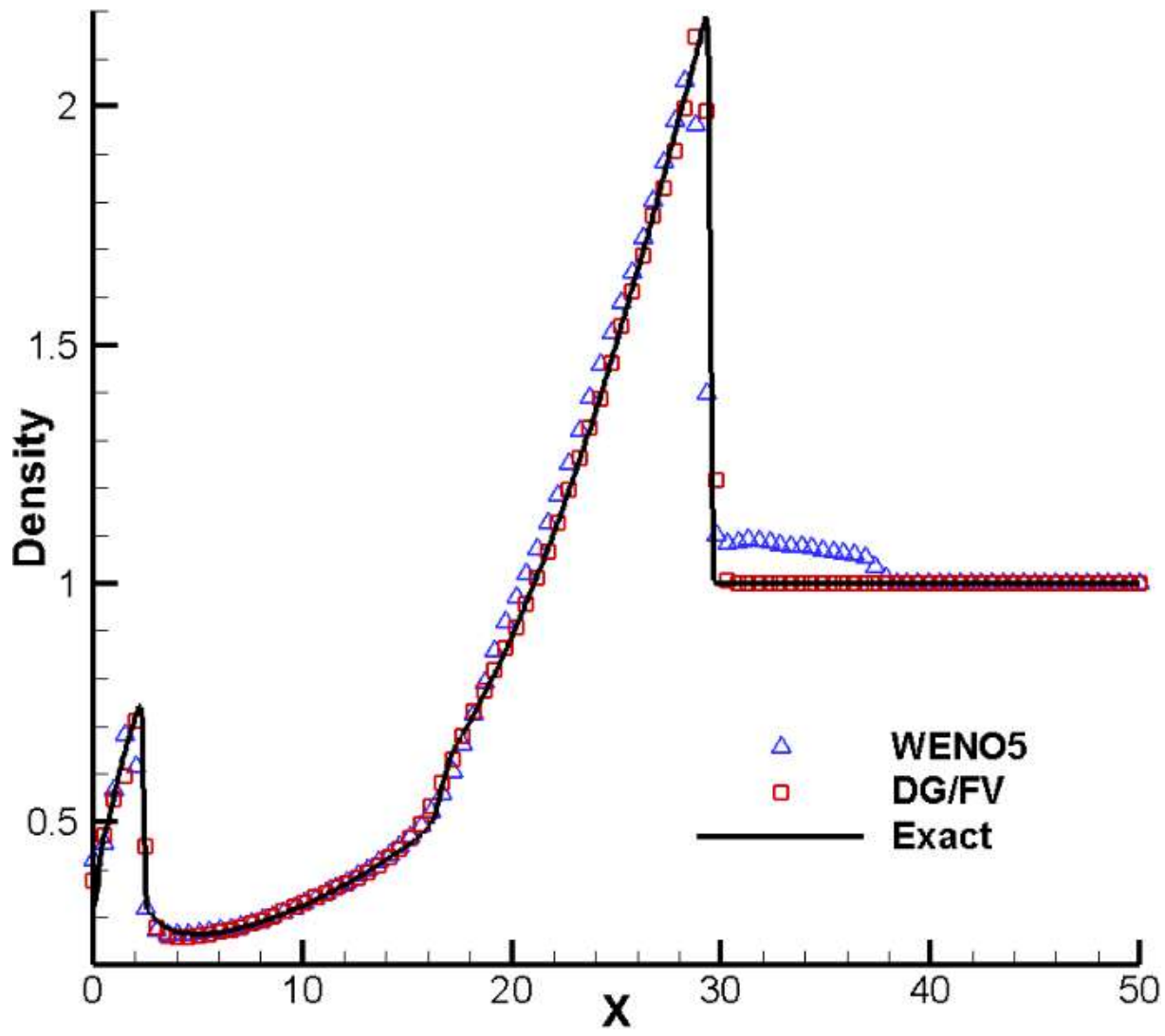

**Fig. 27** Radially symmetric detonation wave problem. The density profile at the cross-section $y=x$ at $t=4$.

### 4.2.7 A multi-species multi-reactions detonation wave

The last example is a 2D detonation wave with five species and two reactions, which has been studied in [57]. The reacting models with five species and two reactions are

$$
\begin{aligned}
&H_2 + O_2 \rightarrow 2OH, \\
&2OH + H_2 \rightarrow 2H_2O.
\end{aligned}
\tag{29}
$$

The species $N_2$ appears as a catalyst. The parameters are $\gamma=1.4$, $T_{ign,1}=2$, $T_{ign,2}=10$, $1/\varepsilon_1=10^5$, $1/\varepsilon_2=2\times10^4$, $q_1=0$, $q_2=0$, $q_3=-20$, $q_4=-100$, $q_5=0$, $W_1=2$, $W_2=32$, $W_3=17$, $W_4=18$ and $W_5=28$. The computation domain is $[0, 150]\times[0, 25]$. The initial conditions are

$$
\begin{aligned}
&(\rho,\ u,\ v,\ p,\ z_1,\ z_2,\ z_3,\ z_4,\ z_5) = \\
&\begin{cases}
(2,\ 10,\ 0,\ 40,\ 0,\ 0,\ 0.17,\ 0.63,\ 0.2), & x \le \xi(y), \\
(1,\ 0,\ 0,\ 1,\ 0.08,\ 0.72,\ 0,\ 0,\ 0.2), & x > \xi(y),
\end{cases}
\end{aligned}
\tag{30}
$$

where

$$
\xi(y) = \begin{cases}
12.5 - |y-12.5|, & |y-12.5| \le 7.5, \\
5, & |y-12.5| > 7.5.
\end{cases}
\tag{31}
$$

The inflow and outflow boundary conditions are adopted on the left and right boundaries, respectively. The solid wall boundary condition is applied on the top and bottom boundaries. Similar cases can also be found for instance in [58] and [59]. Based on our numerical experience, this multi-species multi-reactions test case has even stronger stiffness compared to previous cases with the simplified reaction model, and thus requires higher resolution of the numerical method. In the test case we use the technique of adding the fifth-order reconstruction and using a two-staged BVD selection algorithm, and we use the characteristic variables in the reconstruction procedure. Fig. 28 gives the evolution of the detonation wave at $t=2$, $t=4$, $t=6$ and $t=8$ obtained by the third-order DG/FV scheme on a mesh with $300\times50$ elements. The results presented here get close to the results obtained by the fifth-order WENO using $1500\times250$ elements, see the right figures presented in Fig. 27 in [58]. Fig. 29 plots the computed pressure and density along $y=12.5$ at $t=8$, as well as the reference solution from [58]. The overall agreement is seen between the present method and the WENO method performed on the refined mesh. Moreover, we also conduct a simulation by utilizing the third-order scheme on a refined mesh with $900\times150$ elements. Fig. 30 illustrates the density solutions at $t=1$ and $t=2$, which show delicate flow structures behind the detonation wave. One can also see Fig. 4.5 in [59] for a comparison. Overall, the method has achieved oscillation-less and positivity-preserving solutions on the subcell for stiff multi-species detonations.

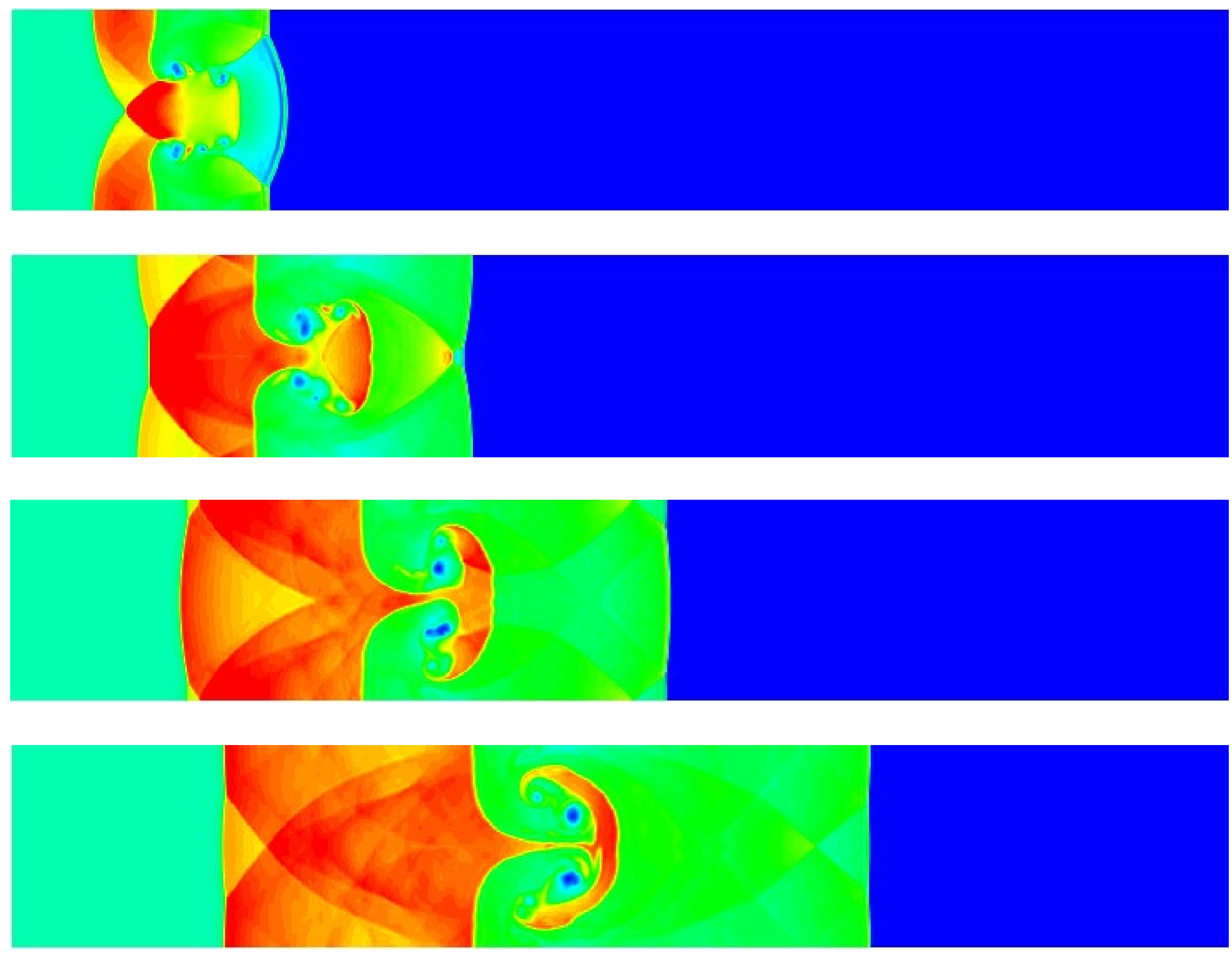

**Fig. 28** Multi-species multi-reactions detonation wave problem. Density counters at $t$=2, $t$=4, $t$=6 and $t$=8 on the 300×50 mesh.

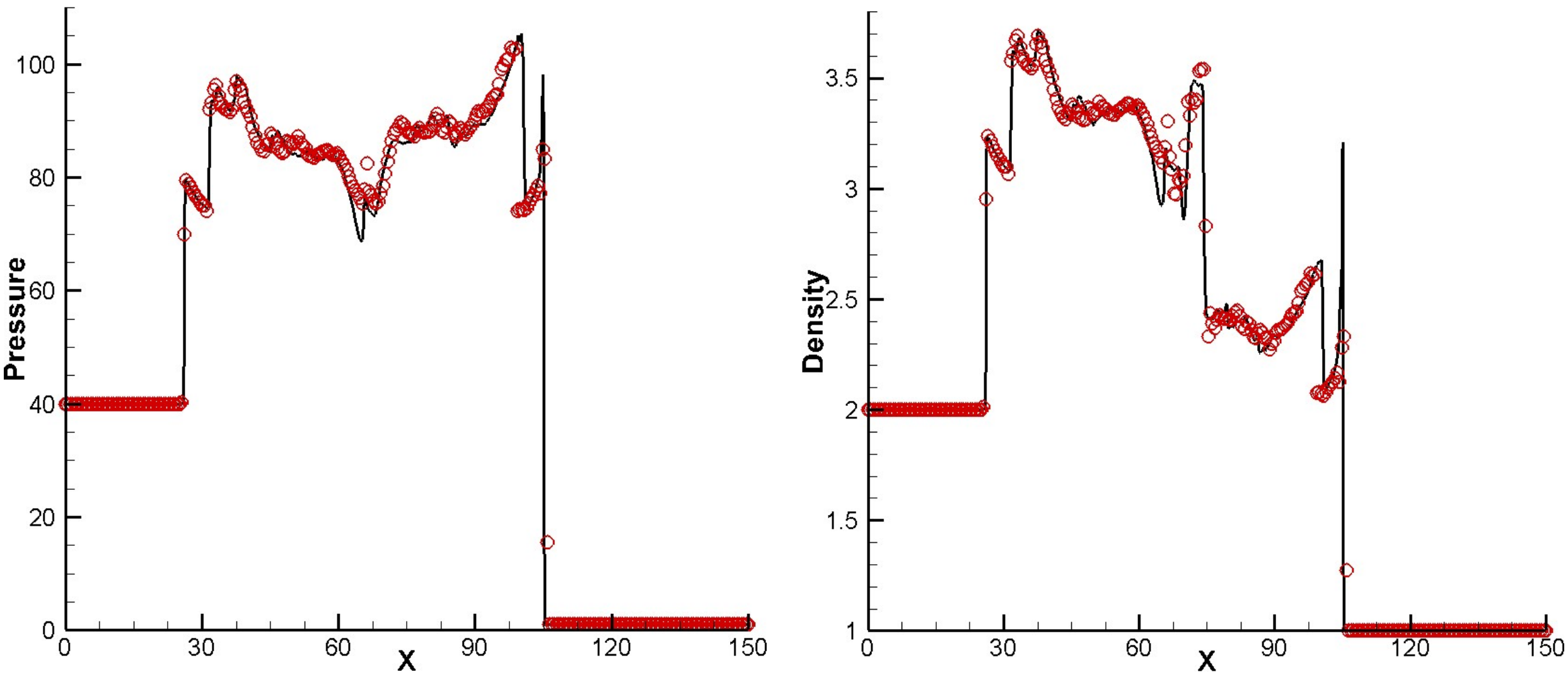

**Fig. 29** Multi-species multi-reactions detonation wave problem. The pressure and density distribution along $y$=12.5 at $t$=8. The black solid lines denote the WENO method performed on the refined mesh from [58]. The red symbols denote the present method performed on the 300×50 mesh.

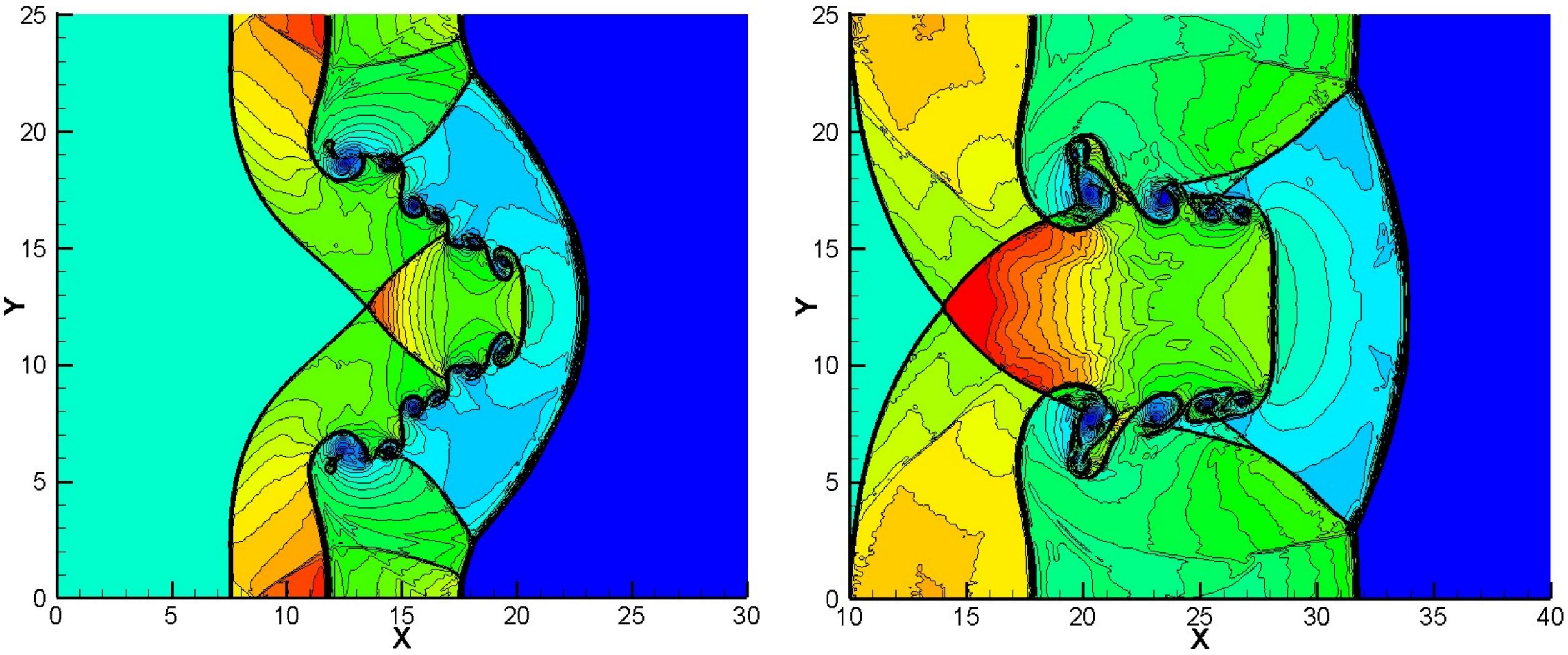

**Fig. 30** Multi-species multi-reactions detonation wave problem. Density solutions with 30 counter lines in the range of [1, 4] at $t$=1 (left) and $t$=2 (right) performed with the 3rd-order DG/FV scheme on the 900×150 mesh.

## 5 Concluding remarks

In this paper, an approach that seeks to preserve the positivity of the density and pressure within the framework of hybrid DG/FV discretization has been presented. Through an analysis of the positivity constraint for the DG/FV discretization, we have designed the positivity-preserving strategy consisting of the *a priori* and *a posteriori* detection/computation procedure for the hybrid DG/FV method. The *a priori* detection acts on the process of determining the troubled cells and the process of the FV computation on the subcell. The information on physics-based admissibility property as well as the information on subcell data reconstruction is explored in the *a priori* detection. The *a priori* computation utilizes the THINC scheme to preserve the boundedness of the reconstruction. On the other hand, the *a posteriori* detection/computation procedure maintains the positivity through the detection on the subcell solution and the implementation of the first-order Godunov scheme to those subcells with extremely complex computational condition. Furthermore, a technique of adaptive reconstruction is suggested to reduce the excessive numerical dissipation in the shock-capturing scheme in order to reproduce the correct position of detonation waves when solving the reactive Euler equations with stiff source term. The resulting method is able to obtain positivity-preserving and oscillation-less numerical solutions at the subgrid level so that the subcell resolution capability of the DG method can be fully exploited. We have tested the designed method on a stringent test suite in which the density or pressure may easily become negative. The numerical results verify that the method is able to obtain the positive subcell solutions and capture the flow details using rather coarse meshes. For stiff detonation waves, the present method resolves the correct position of the detonation front and achieves comparable and even better resolution in comparison with the FV method using the same DOFs and the fifth-order WENO reconstruction scheme.

## Acknowledgments

This research is supported by the National Natural Science Foundation of China under Grants No. 11702015, 11721202 and 92252201. We also acknowledge the support of the Fundamental Research Funds for the Central Universities and the support of Beijing Natural Science Foundation (L248036).

## Appendix: the description of the chemical reaction source term

### A. Simplified model

In this model, the mass fraction $z_k$=$z$. We have

$$p = (\gamma - 1)\rho(e - \frac{1}{2}\boldsymbol{u}\cdot\boldsymbol{u} - zq_0), \tag{A.1}$$

where $\gamma$ is the specific heat ratio, $q_0$ is the chemical heat release. The source term $\omega_k$=$\omega$ that is expressed by

$$\omega = -K(T)\rho z. \tag{A.2}$$

$T$ is the temperature and given by

$$T = \frac{p}{\rho}. \tag{A.3}$$

In the simplified model, the reaction rate $K(T)$ is modeled in two ways. One is through the Arrhenius form

$$K(T) = K_0 e^{-T_{\text{ign}}/T}, \tag{A.4}$$

where $K_0$ is the reaction rate constant and $T_{\text{ign}}$ is the ignition temperature. The other is through the Heaviside form

$$K(T) = -\frac{1}{\xi}H(T - T_{\text{ign}}), \tag{A.5}$$

where $\xi$ represents the reaction time. In Eq. (A.5), we have $H(x)$=1 for $x\geq 0$ and $H(x)$=0 for $x$<0.

In this work, the simplified model is used for test cases from Sec. 4.2.1 to Sec. 4.2.5.

### B. Multi-species and multi-reactions model

In this model, we have

$$p = (\gamma - 1)\rho(e - \frac{1}{2}\boldsymbol{u}\cdot\boldsymbol{u} - \sum_{k=1}^{ns} z_k q_k), \tag{A.6}$$

where $ns$ is the species number and $q_k$ is the chemical heat release of the $k$-th species. There are assumed to be $R$ reactions of the form

$$\upsilon'_{1,r} X_1 + \upsilon'_{2,r} X_2 + \cdots + \upsilon'_{ns,r} X_{ns} \rightarrow \upsilon''_{1,r} X_1 + \upsilon''_{2,r} X_2 + \cdots + \upsilon''_{ns,r} X_{ns}, \quad r = 1, \cdots, R, \tag{A.7}$$

where $\upsilon'$ and $\upsilon''$ are the stoichiometric coefficients of the reactants. The source term is expressed by

$$\omega_k = W_k \sum_{r=1}^{R} \left(\upsilon''_{k,r} - \upsilon'_{k,r}\right) \left[ K_r(T) \prod_{s=1}^{ns} \left( \frac{\rho z_s}{W_s} \right)^{\upsilon'_{s,r}} \right], \quad k = 1, \cdots, ns, \tag{A.8}$$

where $W_k$ is the molar mass of the $k$-th species. The temperature is computed as Eq. (A.3). For reaction $r$, the reaction rate $K_r(T)$ is modeled by the Heaviside form

$$K_r(T) = \frac{1}{\varepsilon_r} H(T - T_{\text{ign},r}), \tag{A.9}$$

where $\varepsilon_r$ and $T_{\text{ign},r}$ are constants for the $r$-th reaction.

In this work, the multi-species and multi-reactions model is used for the test case in Sec. 4.2.6 and Sec.

4.2.7.